\documentclass[preprint,showpacs,preprintnumbers,amsmath,amssymb,nofootinbib]{revtex4}

\usepackage{graphicx}
\usepackage{bm}

\newcommand{\nnn}{\nonumber \\}
\newcommand{\kmax}{k_{\rm max}}

\begin{document}

\preprint{HUPD0910, UT-10-07}

\title{Effects of Large Threshold Corrections in Supersymmetric Type-I Seesaw Model}

\author{Sin Kyu Kang$^1$}
\author{Takuya Morozumi$^2$}
\author{Norimi Yokozaki$^3$}%
\affiliation{%
\\
$^1$ School of Liberal Arts, Seoul National University of Technology, Seoul 139-473, Korea. \\
$^2$ Graduate School of Science, Hiroshima University,
Higashi-Hiroshima, 739-8526, Japan.\\
$^3$ Department of Physics, University of Tokyo,
Tokyo 113-0033, Japan.
}%


\begin{abstract}
We investigate lepton flavor violating (LFV) radiative processes
and the relic abundance of neutralino dark matter in supersymmetric type-I seesaw model.
We carefully derive  threshold corrections to the flavor off-diagonal
elements of slepton mass matrix and up-type Higgs mass squared and find that they can be large in the case of large $B_N^0$.
We examine how the branching ratios of LFV radiative decays and the relic abundance of neutralino dark matter
can be significantly affected by the large threshold corrections.
Soft scalar mass squared parameter of up-type Higgs scalar is also affected by the threshold corrections.
Since the higgsino mass depends on the mass parameter for up-type Higgs, the LFV processes and the relic abundance of
 the neutralino dark matter are correlated with each other.
We show that there are parameter regions where the predictions of the relic abundance of neutralino dark matter are consistent with
WMAP observation and  the branching ratios of LFV radiative decays are predicted to be testable in future experiments.
We find that the masses of scalar supersymmetric particles are not necessarily small so that the branching ratios of LFV decays
can be testable in future experiment, which is distinctive feature of this scenario.
\end{abstract}

\maketitle

\section{Introduction}

Seesaw mechanism has been invented to explain the smallness of observed neutrino masses relative to those of quarks and charged leptons.
Among several varieties of seesaw models, the simplest version is type-I seesaw model which requires the standard model gauge singlet right-handed(RH)
Majorana neutrinos and the existence of a huge mass scale which can be near or similar to the scale of grand unification.
Supersymmetric(SUSY) version of the type-I seesaw model not only inherits this feature but also stabilizes the electroweak scale without fine-tuning,
and provides a natural candidate for a dark matter.
The neutrino Dirac type Yukawa couplings in SUSY type-I seesaw model are flavor off-diagonal, which give rise to neutrino mixing observed by
neutrino oscillation experiments.
Thanks to those Yukawa couplings flavor off-diagonal elements of the slepton mass matrix are induced by radiative corrections even though
we take slepton mass matrix to be diagonal at the high energy scale.
These flavor off-diagonal slepton masses can enhance branching ratios for lepton flavor violating(LFV) decays 
such as
$\tau\rightarrow\mu\gamma$, $\tau\rightarrow e\gamma$ and $\mu\rightarrow e\gamma$
compared to those in non SUSY seesaw models
\cite{Hisano1995,Hisano1996,Casas2001,Masina:2003wt, Arganda:2005ji, Antusch:2006vw, Arganda:2007jw}. 

In our previous work, we have observed that threshold corrections to Higgs bilinear terms mediated by RH sneutrino can affect
the minimization condition for the Higgs potential and thus the fine-tuning may be reduced
when the mass splitting of RH sneutrinos becomes large\cite{Kang:2009pj}.
We have shown that such a large mass splitting can be originated from large value of the B-term for RH sneutrino, \footnote{
The large value of $B_N$, (i.e. $B_N \gg m_{soft}\sim B (\sim \mbox{electorweak scale}$)) can be obtained from the term of superpotential given by \cite{Farzan:2004cm},
\begin{eqnarray*}
 W = \int d^2\theta \lambda X N^2,
\end{eqnarray*}
with $M_R \equiv \lambda \left<X\right>$, $B_N \equiv \left<F_X \right>/\left<X \right>$. In the case that $\left<X\right>\sim 10^{16}\, {\rm GeV}$
and $\left<F_X \right> \sim 10^{21}\,{\rm GeV}^2$ which is the same order of SUSY breaking F-term, $B_N \sim 100\, {\rm TeV}$ is obtained.
}
\begin{eqnarray*}
 \mathcal{L} = - \frac{1}{2} B_N M_R \tilde{N}^{* 2} + h.c. .
\end{eqnarray*}
In addition, we have also shown that those threshold corrections can significantly affect the relic abundance of the neutralino dark matter
in minimal supergravity scenario(mSUGRA). Thus some of the parameter space excluded by WMAP data in SUSY seesaw model without the threshold corrections
can be consistent with WMAP data when we include the threshold corrections.
However, we have not considered the flavor effects generated from the threshold corrections.

In this paper, we show that the threshold corrections can give rise to sizable contributions to lepton flavor violating phenomena,
and extensively discuss how the branching ratios of LFV radiative decays are correlated with the relic abundance of neutralino dark matter.
We believe that this observation is new although there are several literatures where the contributions of $B$-parameter and threshold
corrections to lepton flavor violating phenomena have been studied
\cite{Farzan:2003gn,Giudice2010}.

This paper is organized as follows. In section II, we derive the threshold corrections to Higgs and slepton mass squared parameters by
 using RGEs for corresponding parameters. We also calculate the finite terms which are not included in the approach using RGE method.
 In section III, we investigate how the branching ratios of LFV decays and the relic abundance of neutralino dark matter are affected
 by the threshold corrections and correlated with each other.
 In section IV, we devote to the numerical calculation and present our results.
The concluding remarks will follow in section V.
One-loop RGEs for slepton and Higgs masses including threshold corrections are presented in appendix.

\section{The threshold Corrections}
In SUSY type-I seesaw model, flavor off-diagonal elements of $SU(2)$ slepton mass matrix can arise
from radiative corrections mediated by RH neutrinos and sneutrinos even though the slepton mass matrix is taken to be
flavor-diagonal at the high energy scale such as the GUT scale.
These corrections are evaluated with the help of  RGEs given in Ref.\cite{Hisano1996} under the assumption that soft scalar masses and
scalar trilinear couplings are universal at the high energy scale.
When the mass splittings of RH sneutrinos are large, threshold corrections arisen from integrating out heavy sectors
should be taken into account. As we studied in \cite{Kang:2009pj}, large mass splitting of RH sneutrinos is originated from
large value of $B$-terms for RH sneutrinos.
It turns out from our numerical estimation that those threshold corrections can dominate over the other radiative corrections
to the flavor off-diagonal elements of $SU(2)$ slepton mass matrix.
It is also worthwhile to notice that both radiative corrections and threshold corrections to the flavor off-diagonal elements of slepton mass matrix
can significantly contribute to Higgs mass squared parameters\cite{Farzan:2004cm,Kang:2009pj}.

\begin{figure}[htbp]
\includegraphics[width=7cm]{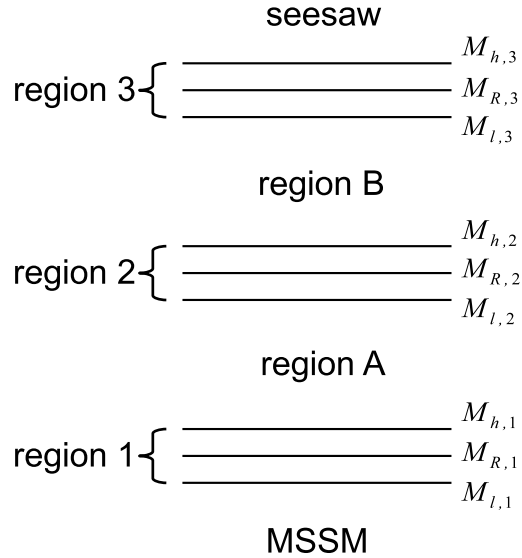}
\caption{Mass hierarchies of RH neutrino and sneutrino masses. $M_{R,i}$ denote RH neutrino masses whereas $M_{h,i}$ and $M_{l,i}$
are heavier and lighter sneutrino masses, respectively}
\label{fig:thresholds}
\end{figure}

Now, let us derive the threshold corrections to slepton and Higgs masses by integrating out
one-loop RGEs in the case that the mass differences among three generations of RH sneutrinos are large.
The superpotential in SUSY type-I seesaw model is given as
\begin{eqnarray}
W = Y_{e,ij} H_1\cdot L_i \bar{E}_j + Y_{\nu,ij} L_i \cdot H_2 \bar{N}_j + \frac{1}{2}M_{R,ij} \bar{N}_i \bar{N}_j - \mu H_1 \cdot H_2 ,
\end{eqnarray}
and soft SUSY breaking terms are written as
\begin{eqnarray}
\mathcal{L}_{soft} =&& -m_{\tilde{E},ij}^2 \tilde{E}_i^\dag \tilde{E}_j - m_{\tilde{L},ij}^2 \tilde{L}_i^\dag \tilde{L}_j
- m_{\tilde{N},ij}^2 \tilde{N}_i^\dag \tilde{N}_j - m_{H_1}^2 H_1^\dag H_1 - m_{H_2}^2 H_2^\dag H_2 \nnn
&& + \left(m^2_{H_1 H_2} H_1 \cdot H_2 + h.c. \right)  - \left(A_{\nu,ij} \tilde{L}_i \cdot H_2 \tilde{N}_j^* + h.c. \right) \nnn
&& - \left(A_{e,ij} H_1 \cdot \tilde{L}_i \tilde{e}_{R,j}^* + h.c. \right)
- \frac{1}{2} \left(B^2_{ij} \tilde{N}_i^* \tilde{N}_j^* + h.c. \right) \ .
\end{eqnarray}
Redefining the chiral superfields, $\bar{N}_i$, we can take $M_{R,ij}$ to be diagonal as
\begin{eqnarray}
 M_{R,ij} = M_{R,i} \delta_{ij},
\end{eqnarray}
where $M_{R,i}$ are real, positive and
assumed to be hierarchical, i.e. $M_{R,1} \ll M_{R,2} \ll M_{R,3}$.
Then, in the case of $|B^2_{ij}| \ll |M_{R,i}^2-M_{R,j}^2|$,
the mass eigenvalues of RH sneutrinos are approximately given as
\begin{eqnarray}
 M_{h,i}^2 = m_{\tilde{N},ii}^2 + M_{R,i}^2 + |B^2_{ii}|, \nnn
 M_{l,i}^2 = m_{\tilde{N},ii}^2 + M_{R,i}^2 - |B^2_{ii}|, \label{eq:hsn_threshold}
\end{eqnarray}
Here, the mass eigenstates, $\tilde{N}_{h,i}$ and $\tilde{N}_{l,i}$ are given by
\begin{eqnarray}
 \tilde{N}_i = \frac{1}{\sqrt{2}} e^{i \Phi_i/2} (\tilde{N}_{h,i} + i \tilde{N}_{l,i}), \nnn
 \tilde{N}_i^{*} = \frac{1}{\sqrt{2}} e^{-i \Phi_i/2} (\tilde{N}_{h,i} - i \tilde{N}_{l,i}), \label{eq:msplit}
\end{eqnarray}
where
\begin{eqnarray}
 \Phi_i = {\rm arg}(B^2_{ii}).
\end{eqnarray}
The hierarchical structure of RH neutrino and sneutrino masses is presented in Fig.\ref{fig:thresholds}.
The flavor off-diagonal part of the mass terms for RH sneutrinos is given as
\begin{eqnarray}
 \mathcal{L} &=&
- \frac{1}{2} \sum_{i\neq j} \left(m_{h,ij}^2 \tilde{N}_{h,i} \tilde{N}_{h,j} +  m_{l,ij}^2 \tilde{N}_{l,i} \tilde{N}_{l,j}
+ m_{hl,ij}^2 \tilde{N}_{h,i} \tilde{N}_{l,j} + m_{lh,ij}^2 \tilde{N}_{l,i} \tilde{N}_{h,j} \right) ,
\end{eqnarray}
where
\begin{eqnarray}
 m_{h,ij}^2 &=& {\rm Re}(m_{\tilde{N},ij}^2 e^{-i(\Phi_i - \Phi_j)/2}) + {\rm Re}(B^2_{ij} e^{-i(\Phi_i+\Phi_j)/2}) ,\nnn
 m_{l,ij}^2 &=& {\rm Re}(m_{\tilde{N},ij}^2 e^{-i(\Phi_i - \Phi_j)/2}) - {\rm Re}(B^2_{ij} e^{-i(\Phi_i+\Phi_j)/2}) ,\nnn
 m_{hl,ij}^2 &=& -{\rm Im}(m_{\tilde{N},ij}^2 e^{-i(\Phi_i - \Phi_j)/2}) + {\rm Im}(B^2_{ij} e^{-i(\Phi_i+\Phi_j)/2}) ,\nnn
 m_{lh,ij}^2 &=& {\rm Im}(m_{\tilde{N},ij}^2 e^{-i(\Phi_i - \Phi_j)/2}) + {\rm Im}(B^2_{ij} e^{-i(\Phi_i+\Phi_j)/2}) ,\nnn
&=& m_{hl,ji}^2 .
\end{eqnarray}

Let us calculate radiative corrections and threshold corrections to the slepton and Higgs mass squared parameters.
We note that there exist two types of contributions, one is the logarithmic contributions which can be obtained from the RGEs
and the other is finite contributions which are calculated in this paper by the diagrammatic approach.

\subsection{Leading log contributions }
The leading log contributions to slepton  and Higgs mass squared parameters can be obtained from RGEs.
Let us first consider RGEs for slepton masses and up-type Higgs mass without
taking into account the mass threshold contributions of RH sneutrinos.
The explicit expressions of the RGEs arisen from only the contributions due to the
 RH neutrinos and sneutrinos are given by for slepton masses
\cite{Hisano1995,Hisano1996}
\begin{eqnarray}
 (16\pi^2)Q\frac{d m_{\tilde{L},ij}^2}{d Q} &=& 2 (Y_{\nu}^* m_{\tilde{N}}^2 Y_{\nu}^T)_{ij} + 2 (A_{\nu}^* A_{\nu}^T)_{ij}
+ (m_{\tilde{L}}^2 Y_{\nu}^* Y_{\nu}^T)_{ij} + (Y_{\nu}^* Y_{\nu}^T m_{\tilde{L}}^2)_{ij} \nonumber \\
&+& 2 m_{H_2}^2 (Y_{\nu}^* Y_{\nu}^T)_{ij}, \label{slep1}
\end{eqnarray}
and for up-type Higgs mass\cite{Casas2001},
\begin{eqnarray}
 (16\pi^2)Q\frac{d m_{H_2}^2}{d Q} &=& 2 {\rm Tr}(Y_{\nu}^* m_{\tilde{N}}^2 Y_{\nu}^T) + 2 {\rm Tr}(Y_{\nu}^T m_{\tilde{L}}^2 Y_{\nu}^*)
+ 2 m_{H_2}^2 {\rm Tr}(Y_{\nu}^* Y_{\nu}^T) \nonumber \\
&+& 2 {\rm Tr}(A_\nu^* A_\nu^T). \label{up-higg1}
\end{eqnarray}
We assume that soft SUSY breaking parameters are universal at the GUT scale, so that the following relations are valid,
\begin{eqnarray}
 m_{\tilde{L},ij}^2 = m_{\tilde{N},ij}^2 = m_0^2 \delta_{ij},\  m_{H_2}^2=m_0^2,\ A_{\nu,ij} = A_0 Y_{\nu,ij}, \ B_{ij}^2 = B_N^0 M_{R,i} \delta_{ij}.
\label{eq:universal_masses}
\end{eqnarray}
By integrating the above RGEs (\ref{slep1},\ref{up-higg1}) between $M_{R,k} \le Q \le M_{GUT}$, we obtain flavor off-diagonal slepton masses and the radiative corrections to
up-type Higgs masses, and their approximated expressions are given by
\begin{eqnarray}
\delta m_{\tilde{L},ij}^2 \approx - \sum_k \frac{Y_{\nu,ik}^* Y_{\nu,kj}^T}{8\pi^2} \left(A_0^2 +3 m_0^2\right) \ln \frac{M_{GUT}}{M_{R,k}}, \label{eq:run1}
\end{eqnarray}
and
\begin{eqnarray}
 \delta m_{H_2}^2 \approx - \sum_{i,k} \frac{Y_{\nu,ik}^* Y_{\nu,ki}^T}{8\pi^2} \left(A_0^2 +3 m_0^2\right) \ln \frac{M_{GUT}}{M_{R,k}}. \label{eq:run2}
\end{eqnarray}
Here, we have used the fact that when $Q^2 < M_{R,k}^2$ the chiral superfield $\bar{N}_k$ decouples, and thus $Y_{\nu,ik}$ is set to zero.

Next, let us derive the corrections
to the above relations (\ref{eq:run1},\ref{eq:run2}) originated from RH sneutrino mass thresholds shown in Fig.\ref{fig:thresholds}.
We split a complex scalar $\tilde{N}_i$ to two real scalars $\tilde{N}_{h,i}$ and $\tilde{N}_{l,i}$ as in Eq.(\ref{eq:msplit}) and then
re-derive RGEs by using the method given in Ref.\cite{Luo2003}. The complete results are presented in Appendix A.
The terms of the RGEs which contribute to the derivation of the threshold corrections are written as
\begin{eqnarray}
 (16\pi^2) Q \frac{d m_{\tilde{L},ij}^2}{d Q} &\ni&
 2 \sum_{k} Y_{\nu,ik}^* Y_{\nu,kj}^T M_{R,k}^2
\left[ \theta(Q^2-M_{h,k}^2)+\theta(Q^2-M_{l,k}^2)-2\theta(Q^2-M_{R,k}^2)\right] \nnn
&+& \sum_k \left[Y_{\nu,ik}^* A_{\nu,kj}^T e^{-i\Phi_k} + A_{\nu,ik}^* Y_{\nu,kj}^T e^{i\Phi_k}\right] M_{R,k} \nnn
&& \times \left[\theta(Q^2-M_{h,k}^2) - \theta(Q^2-M_{l,k}^2)\right] \nnn
&+& \sum_{k}\left[ Y_{\nu,ik}^* |B^2_{kk}|Y_{\nu,kj}^T
+ \sum_{k'=1}^{k-1} \left(Y_{\nu,ik'}^* B_{k'k}^2 e^{-i\Phi_k} Y_{\nu,kj}^T
+ Y_{\nu,ik}^* B_{kk'}^{2*} e^{i\Phi_k} Y_{\nu,k'j}^T \right)
\right] \nnn
&& \times \left[\theta(Q^2-M_{h,k}^2)-\theta(Q^2-M_{l,k}^2)\right],
\label{eq:RGE_slepton}
\end{eqnarray}
and
\begin{eqnarray}
 (16\pi^2)Q\frac{d m_{H_2}^2}{d Q}
&\ni& 2 \sum_{i,k} Y_{\nu,ik}^* Y_{\nu,ki}^T M_{R,k}^2 \left[\theta(Q^2-M_{h,k}^2)+\theta(Q^2-M_{l,k}^2)
- 2 \theta(Q^2-M_{R,k}^2)\right] \nnn
&+& 2 \sum_{i,k} {\rm Re}(Y_{\nu,ik}^* A_{\nu,ki}^T e^{-i\Phi_k}) M_{R,k}
\left[\theta(Q^2-M_{h,k}^2)-\theta(Q^2-M_{l,k}^2)\right] \nnn
&+& \sum_{i,k}\left[ Y_{\nu,ik}^* |B^2_{kk}| Y_{\nu,ki}^T
+ 2 \sum_{j=1}^{k-1} {\rm Re}\left(Y_{\nu,ij}^* B_{jk}^2 e^{-i\Phi_k} Y_{\nu,ki}^T \right)
\right] \nnn
&& \times \left[\theta(Q^2-M_{h,k}^2)-
\theta(Q^2-M_{l,k}^2)\right] . \label{eq:RGE_higgs}
\end{eqnarray}
Note that the terms of RGEs given in Eq.(\ref{eq:RGE_slepton}) and (\ref{eq:RGE_higgs})
have been derived in the diagonal basis of Majorana mass matrix
$M_{R}$ by following the method appeared in \cite{Sasaki:1986jv}.
%
%
Integrating the RGEs for the slepton masses given in Appendix A for the region of energy scale, $M_{l,k}^2 \le Q^2 \le M_{h,k}^2$,
we obtain the threshold corrections as follows:
\begin{eqnarray}
\delta_{th}^k m_{\tilde{L},ij}^2 &\equiv& m_{\tilde{L},ij}^2(M_{l,k}^2)- m_{\tilde{L},ij}^2(M_{h,k}^2) \nnn
&=& \delta_{th1}^k m_{\tilde{L},ij}^2 + \delta_{th2}^k m_{\tilde{L},ij}^2 ,
\end{eqnarray}
where
\begin{eqnarray}
\delta_{th1}^k m_{\tilde{L},ij}^2 &=& \frac{1}{16\pi^2}
\left(2Y_{\nu,ik}^* Y_{\nu,kj}^T m_{\tilde{N},kk}^2 +
Y_{\nu,ik}^* A_{\nu,kj}^T \frac{B^{2*}_{kk}}{M_{R,k}}  + {A}_{\nu,ik}^* Y_{\nu,kj}^T \frac{B_{kk}^{2}}{M_{R,k}} \right),
\label{eq:thresha}
\end{eqnarray}
\begin{eqnarray}
\delta_{th2}^k m_{\tilde{L},ij}^2 &=& \frac{1}{16\pi^2}\sum_{m=1}^{k-1}
\left(B_{mk}^{2} B_{kk}^{2*} Y_{\nu,im}^* Y_{\nu,kj}^T + B_{mk}^{2*} B_{kk}^2 Y_{\nu,ik}^* Y_{\nu,mj}^T \right) \frac{1}{M_{R,k}^2}.
\label{eq:threshb}
\end{eqnarray}
The terms which contain both $A_\nu$ and $B^2$ in Eq.(\ref{eq:thresha}) agree with Ref.\cite{Giudice2010} whereas disagree with Ref.\cite{Farzan:2003gn}.
The authors of Ref.\cite{Giudice2010} derived the threshold corrections by using the method of analytic
continuation into superspace\cite{Giudice1998,Arkani-Hamed1998,Matsuura2007}
and the author of Ref.\cite{Farzan:2003gn} derived the corrections by diagrammatic calculation.
In the derivation of the above equations, we have used the following relations,
\begin{eqnarray}
 \ln \frac{M_{h,k}^2}{M_{R,k}^2} = \frac{|B^2_{kk}|}{M_{R,k}^2} + \frac{m_{\tilde{N},kk}^2}{M_{R,k}^2} -\frac{1}{2}\frac{|B^2_{kk}|^2}{M_{R,k}^4}
+ \mathcal{O}(M_{R,k}^{-3}) , \ \ln\frac{M_{h,k}^2}{M_{l,k}^2} = 2 \frac{|B_{kk}^2|}{M_{R,k}^2} + \mathcal{O}(M_{R,k}^{-3}).
\end{eqnarray}
Imposing universal condition for the soft SUSY breaking parameters at the GUT scale, $\delta_{th}^k m_{\tilde{L},ij}^2$ is approximately written as
\begin{eqnarray}
\delta_{th}^k m_{\tilde{L},ij}^2 &\approx& \frac{1}{8\pi^2} Y_{\nu,ik}^* Y_{\nu,kj}^T (m_0^2 + A_0 B_N^0) \label{eq:th_mlsq}.
\end{eqnarray}
We note that since $B^2_{ij}(i\neq j)$ are radiatively generated at one-loop level,
$\delta_{th2}^k m_{\tilde{L},ij}^2$ are two-loop contributions.
To estimate how large it is,
we solve RGE for $B^2_{ij}$ given in Eq.(\ref{eq:RGEs_B}), and then
obtain its value around at $Q=M_{R,3}$,
\begin{eqnarray}
B^2_{mk}(M_{R3}) &=&
%
%
B_N^0M_{R,k}\left[\delta_{m k}-
\frac{(Y^T_{\nu} Y_\nu^\ast)_{mk}(1+\delta_{mk})}{8 \pi^2}
\ln \frac{M_{GUT}}{M_{R,3}} \right] \nonumber \\
&-&A_0 M_{R,k}
\frac{(Y^T_{\nu} Y_\nu^\ast)_{mk}(1+\delta_{mk})}{4 \pi^2}
\ln \frac{M_{GUT}}{M_{R,3}}
,
\label{eq:B}
\end{eqnarray}
where $m \leq k$.
We notice that the term proportional to $B^0_N$ in the RH side of Eq.(\ref{eq:B}) can be diagonalized by
changing the basis of $M_{R}$ into the diagonal basis at $Q=M_{R,3}$, but the term proportional to
$A_0$ can not be simultaneously diagonalized as proven in appendix C.
Then, the contributions to $\delta_{th2}^3 m_{\tilde{L},ij}^2$
are approximately given by
\begin{eqnarray}
-\frac{B_{N}^0}{ 4 \pi^2}
 \frac{A_0}{16{\pi^2}}\left[\sum_{m=1}^{2}
\left(Y_{\nu,im}^* (Y_{\nu}^T Y_{\nu}^*)_{m3}
Y_{\nu,3j}^T + Y_{\nu,i3}^*
(Y_{\nu}^T Y_{\nu}^*)_{3m}
Y_{\nu,mj}^T \right)\right] \ln
\frac{M_{GUT}}{M_{R,3}},
\end{eqnarray}
which turns out to be much smaller than $\delta_{th1}^3 m_{\tilde{L},ij}^2$, so that we can neglect these contributions.

Similarly, integrating Eq.(\ref{eq:RGE_higgs}) for the range of the scale, $M_{l,k}^2 \le Q^2 \le M_{h,k}^2$,
we obtain the threshold corrections to
the Higgs mass $m_{H_2}^2$ given by,
\begin{eqnarray}
 \delta^k_{th} m_{H_2}^2 \approx  \sum_i \frac{1}{8\pi^2} Y_{\nu,ik}^* Y_{\nu,ki}^T (m_0^2 + A_0 B_N^0).
\end{eqnarray}

\subsection{Scheme dependent finite terms}
\begin{figure}[htbp]
\includegraphics[width=15cm]{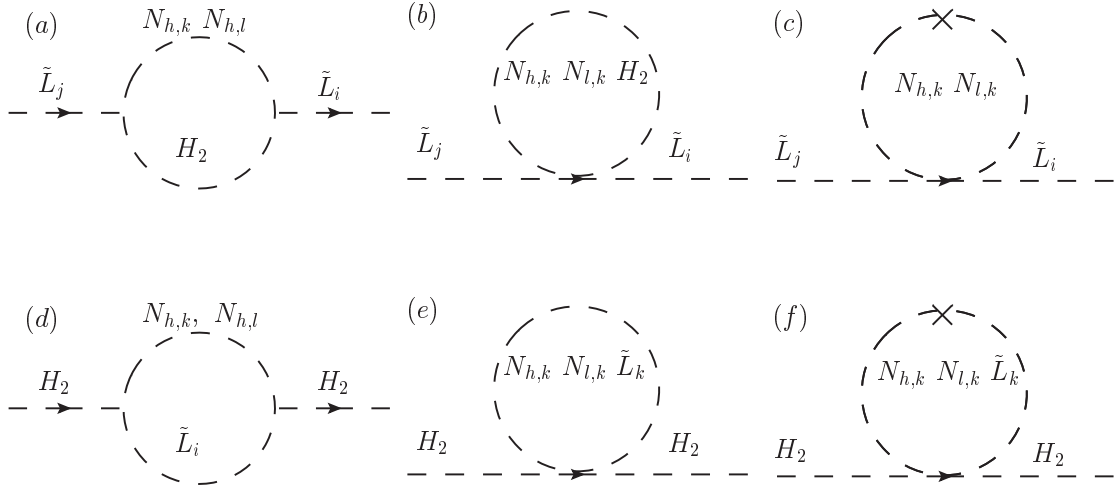}
\caption{Contributions from scalar degrees of freedom}
\label{fig:diagrams}
\end{figure}

\begin{figure}[htbp]
\includegraphics[width=15cm]{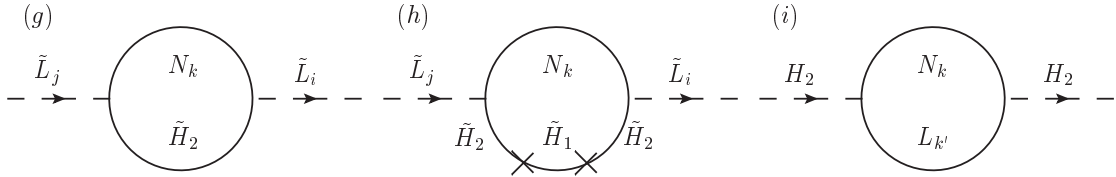}
\caption{Contributions from fermionic degrees of freedom}
\label{fig:diagrams2}
\end{figure}

There are also finite terms which are renormalization scheme dependent. We obtain them by calculating corresponding Feynman diagrams and
estimate how large they are. We used $\overline{\rm MS}$ ($\overline{\rm DR}$) scheme to subtract divergences. It turns out that these terms are small compared to the logarithmic contributions. To derive the
scheme dependent terms,
we first consider the divergent diagrams which arise from RH neutrino-sneutrino Yukawa interactions.
The Feynman diagrams relevant to our purpose are
shown in Figs.\ref{fig:diagrams},\ref{fig:diagrams2}.
If we insert more flavor off-diagonal slepton masses into those diagrams, they can lead to only finite corrections proportional to
the mass parameters such as $B_{ij}^2$, $m_{\tilde{L},ij}^2$ and $m_{\tilde{N},ij}^2$ with $i\neq j$, which are at most at two-loop level because
the mass parameters are radiatively generated at one-loop level under the assumption that soft masses are universal at the GUT scale.

The one-loop corrections to $m_{\tilde{L},ij}^2$ are obtained by calculating the diagrams (a)-(c)(g)(h) in Figs. 2,3.
The corrections are composed of two contributions arisen from scalar loops and fermion loops,
\begin{eqnarray}
\Sigma_{\tilde{L},ij}=\Sigma^{s}_{\tilde{L},ij}+\Sigma^{f}_{\tilde{L},ij}.
\end{eqnarray}
The scalar loop contributions are given by
\begin{eqnarray}
-i\Sigma^{s}_{\tilde{L},ij} &=& \frac{1}{2} \left(A_{\nu,ik}^* A_{\nu,kj}^T + Y_{\nu,ik}^* Y_{\nu,kj}^T M_{R,k}^2 \right)\left[I_2(m_{22}^2, M_{h,k}^2)+I_2(m_{22}^2, M_{l,k}^2)\right] \nonumber \\
&+& \frac{1}{2}(Y_{\nu,ik}^* A^T_{\nu, kj}e^{-i\Phi_k} + A_{\nu, ik}^* Y_{\nu,kj}^T e^{i\Phi_k})M_{R,k}\left[I_2(m_{22}^2, M_{h,k}^2)-
I_2(m_{22}^2, M_{l,k}^2)\right] \nonumber \\
&+& \frac{1}{2}Y_{\nu,ik}^* Y_{\nu,kj}^T[2I_1(m_{22}^2) + (1+{|\mu|^2}/{M_{h,k}^2})I_1(M_{h,k}^2)
+ (1+|\mu|^2/M_{l,k}^2)I_1(M_{l,k}^2)] \nonumber \\
&+& \frac{1}{2}Y_{\nu,ik}^* Y_{\nu,k'j}^T e^{\frac{i}{2}(\Phi_k-\Phi_k')}\Bigl[ m_{h,kk'}^2 I_2(M_{h,k}^2,M_{h,k'}^2) + m_{l,kk'}^2 I_2(M_{l,k}^2,M_{l,k'}^2) \nonumber \\
&&-i m_{hl,kk'}^2 I_2(M_{h,k}^2, M_{l,k'}^2) + im_{hl,k'k}^2 I_2(M_{h,k'}^2, M_{l,k}^2)\Bigr]+ i \frac{p^2}{32 \pi^2} Y^\ast_{\nu,ik} Y^T_{\nu,kj},
\label{eq:dia_slep_scalar}
\end{eqnarray}
where $m_{22}^2 = |\mu|^2 + m_{H_2}^2$. The last momentum dependent term of Eq.(\ref{eq:dia_slep_scalar}) is obtained by
expanding $-i\Sigma^{s}_{\tilde{L},ij}$
with respect to external momenta $\frac{p^2}{M_{R,k}^2}$
and by keeping the term
which remains in the limit of large $M_{R,k}$.
The loop functions $I_1$ and $I_2$ are given as,
\begin{eqnarray}
 I_1(m^2) &=& Q^{4-d} \int\frac{d^d k}{(2\pi)^d}\frac{1}{k^2-m^2} \nonumber \\
&=& i\frac{m^2}{16\pi^2}\left[\bar{\epsilon}^{-1}+1-\ln\frac{m^2}{Q^2}\right],  \\
 I_2(m_1^2, m_2^2) &=& Q^{4-d} \int \frac{d^d k}{(2\pi)^d} \frac{1}{k^2 - m_1^2} \frac{1}{k^2-m_2^2} \nonumber \\
&=& \frac{i}{16\pi^2}\left[\bar{\epsilon}^{-1}+1+\frac{m_1^2}{m_2^2-m_1^2}\ln \frac{m_1^2}{Q^2}
- \frac{m_2^2}{m_2^2-m_1^2}\ln \frac{m_2^2}{Q^2}
\right] .
\end{eqnarray}

The fermion loop contributions are calculated with the fermion loop diagrams by keeping
external momenta assumed to be small  compared to $M_{R,k}$ so as to derive the contributions from
wave function renormalization.
The results are expressed as
\begin{eqnarray}
 -i\Sigma^f_{\tilde{L},ij} &=& -2 \sum_{k=1}^3Y_{\nu,ik}^* Y_{\nu,kj}^T
 \left( 1+\frac{|\mu|^2}{M_{R,k}^2}\right)I_1(M_{R,k}^2) -i \Sigma_{\tilde{L},ij}^p
\label{eq:dia_slep_fer},
\end{eqnarray}
where $\Sigma_{\tilde{L},ij}^p$ is momentum dependent part given as
\begin{eqnarray}
 -i\Sigma_{\tilde{L},ij}^p &=& \sum_{k=1}^3 \frac{i}{16\pi^2} Y_{\nu,ik}^* Y_{\nu,kj}^T p^2\left(\bar{\epsilon}^{-1}+\frac{1}{2}-\ln\frac{M_{R,k}^2}{Q^2}
 \right) +\mathcal{O}(M_{R,k}^{-1}).
\label{eq:dia_slep_mom}
\end{eqnarray}
The explicit form of $-i\Sigma_{\tilde{L},ij}$ is shown in Appendix B.
%
Then, the quadratic parts of effective Lagrangian for sleptons $\tilde{L}_i$ can be written
in $\overline{{\rm MS}}$ $(\overline{\rm DR})$ scheme as follows,
\begin{eqnarray}
 \mathcal{L}^{k_{\rm max}}_{\rm eff} &=&z^{(\kmax)}_{\tilde{L},ij}(Q^2) \partial_\mu \tilde{L}_{i}^* \partial^\mu \tilde{L}_j \nnn
&-& \left({m^{2(\kmax)}_{\tilde{L},ij}}(Q^2) +
\delta {m^{2(\kmax)}_{\tilde{L},ij}}(Q^2) +
{\delta^{(\kmax)}_{SD} m^2}_{\tilde{L},ij} +
{\delta^{(\kmax)}_{SI} m^2}_{\tilde{L},ij}
\right)\tilde{L}_i^* \tilde{L}_j, \label{eq:effective}
\end{eqnarray}
where the contributions of the loop diagrams
mediated by heavy neutrino superfields, $N_1 \sim N_{\kmax}$ are included and
the parameters are given by
\begin{eqnarray}
z^{(\kmax)}_{\tilde{L},ij}(Q^2)&=&\delta_{ij} -
\sum_{k=1}^{k_{\rm max}}\frac{1}{16\pi^2}
Y_{\nu,ik}^* Y_{\nu,kj}^T
(\ln\frac{M_{R,k}^2}{Q^2}-1), \\
 \delta m_{\tilde{L},ij}^{2(\kmax)}(Q^2) &=& \frac{1}{16\pi^2} \biggl[
\sum_{k=1}^{\kmax} (A_{\nu,ik}^* A_{\nu,kj}^T +
 Y_{\nu,ik}^* Y_{\nu,kj}^T m_{H_2}^2 +
Y_{\nu,ik}^* Y_{\nu,kj}^T m_{\tilde{N},kk}^2)
\ln\frac{M_{R,k}^2}{Q^2} \nnn
&+& \sum_{k\neq k',1}^{\kmax} Y_{\nu,ik}^* Y_{\nu,k'j}^T
m_{\tilde{N},kk'}^2
\ln \frac{{\rm max}(M_{R,k}^2, M_{R,k'}^2)}{Q^2} \biggr] , \\
{\delta^{(\kmax)}_{SD} m^2}_{\tilde{L},ij} &=&
-\frac{1}{16\pi^2} \sum_{k=1}^{\kmax}
(A_{\nu,ik}^* A_{\nu,kj}^T + Y_{\nu,ik}^*
Y_{\nu,kj}^T m_{H_2}^2+ \sum_{k'=1}^{\kmax} Y_{\nu,ik}^* Y_{\nu,k'j}^T
 m_{\tilde{N},kk'}^2) , \\
{\delta^{(\kmax)}_{SI} m^2}_{\tilde{L},ij} &=&
\frac{1}{16\pi^2} \sum_{k=1}^{\kmax}
\left[2 Y_{\nu,ik}^* Y_{\nu,kj}^T m_{\tilde{N},kk}^2 + (Y_{\nu,ik}^* A_{\nu,kj}^T B_{kk}^{2*} + A_{\nu,ik}^* Y_{\nu,kj}^T B_{kk}^2)\frac{1}{M_{R,k}}
\right] .
\end{eqnarray}
Note that $\delta_{SD} m_{\tilde{L},ij}^2$ and
$\delta_{SI} m_{\tilde{L},ij}^2$  are the scheme dependent and independent finite terms, respectively.
One can normalize the kinetic term in canonical form by replacing
$\tilde{L}_i$ as,
\begin{eqnarray}
\sqrt{z_{\tilde{L}}}^{(\kmax)}_{nj} \tilde{L}_j  \to \tilde{L}_n.
\end{eqnarray}
With the replacement, the effective Lagrangian is written as,
\begin{eqnarray}
 \mathcal{L}^{k_{\rm max}}_{\rm eff} &=& \partial_\mu \tilde{L}_{i}^* \partial^\mu \tilde{L}_i \nnn
&-& (\frac{1}{\sqrt{z_{\tilde{L}}}^{(\kmax)}}
{m^{2(\kmax)}_{\tilde{L}}}\frac{1}{\sqrt{z_{\tilde{L}}}^{(\kmax)}}
+\delta {m^{2(\kmax)}_{\tilde{L}}} +
\delta^{(\kmax)}_{SD} m^2_{\tilde{L}} +
\delta^{(\kmax)}_{SI} m^2_{\tilde{L}}
)_{ij}\tilde{L}_i^* \tilde{L}_j. \label{eq:effective_N}
\end{eqnarray}
Now let us apply the matching condition at the scale $Q=M_{R,\kmax}$
to two effective theories, one of which contains the
superfields, $N_1 \sim N_{\kmax}$,
and the other contains only $N_1 \sim N_{\kmax-1}$.
The effective Lagrangian with the $\kmax$ active superfields
should be the same as the one with  the $\kmax-1$ active superfields
at the matching scale;
\begin{eqnarray}
\mathcal{L}^{(\kmax)}_{eff}(Q=M_{R,\kmax})=\mathcal{L}^{(\kmax-1)}_{eff}(Q=M_{R,\kmax}).
\end{eqnarray}
Then we can derive threshold corrections for the soft breaking mass
for sleptons and for the Higgs particle.
Applying the matching condition at $Q^2=M_{R,\kmax}^2$,
we obtain the relation,
\begin{eqnarray}
m^{2(\kmax-1)}_{\tilde{L}}(M_{R,\kmax})
&=&
 \sqrt{z_{\tilde{L}}}^{(\kmax-1)}\frac{1}{\sqrt{z_{\tilde{L}}}^{(\kmax)}}
{m^{2(\kmax)}_{\tilde{L}}(M_{R,\kmax})}\frac{1}{\sqrt{z_{\tilde{L}}}^{(\kmax)}}
{\sqrt{z_{\tilde{L}}}^{(\kmax-1)}}  \nnn
&+&
\delta^{(\kmax)}_{SD} m^2_{\tilde{L}}-\delta^{(\kmax-1)}_{SD}
m^2_{\tilde{L}} \nnn
&+&
\delta^{(\kmax)}_{SI} m^2_{\tilde{L}}-\delta^{(\kmax-1)}_{SI}
m^2_{\tilde{L}}. \label{thres1}
\end{eqnarray}
Here, we have used the relation,
\begin{eqnarray}
\delta m^{2(\kmax)}_{\tilde{L}}(M_{R,\kmax})=\delta
m^{2(\kmax-1)}_{\tilde{L}}(M_{R,\kmax}).
\end{eqnarray}
Inserting the wave function renormalizations given as,
\begin{eqnarray}
\sqrt{z_{\tilde L}}_{nj}^{(\kmax)}&=&\delta_{nj} -
\sum_{k=1}^{\kmax} \frac{1}{32 \pi^2}
Y_{\nu,nk}^* Y_{\nu,kj}^T (\ln\frac{M_{R,k}^2}{Q^2}-1), \\
\frac{1}{\sqrt{z_{\tilde L}}^{(\kmax)}}\Biggr|_{in}&=&
\delta_{in} +\sum_{k=1}^{\kmax} \frac{1}{32 \pi^2}
Y_{\nu,ik}^\ast Y_{\nu,kn}^T(\ln\frac{M_{R,k}^2}{Q^2}-1), \\
\sqrt{z_{\tilde L}}^{(\kmax)}
\frac{1}{\sqrt{z_{\tilde L}}^{(\kmax-1)}}
\Biggr|_{ij}&=&
\delta_{ij}+\frac{1}{32 \pi^2} Y^\ast_{\nu,i\kmax}
Y^T_{\nu,\kmax j}(\log \frac{M_{R,\kmax}^2}{Q^2}-1),
\end{eqnarray}
into Eq.(\ref{thres1}), we finally obtain
\begin{eqnarray}
m^{2(\kmax-1)}_{\tilde{L},ij}(M_{R,\kmax})
&=&
m^{2(\kmax)}_{\tilde{L},ij}(M_{R,\kmax})
- \frac{1}{32 \pi^2}
\left[Y^\ast_{\nu,i \kmax} Y^T_{\nu,\kmax n}
m^{2(\kmax)}_{\tilde{L},nj}(M_{R,\kmax}) \right. \nnn
&+& \left. m^{2(\kmax)}_{\tilde{L},in}(M_{R,\kmax})
Y^\ast_{\nu,n \kmax} Y^T_{\nu,\kmax j}\right]
-
\frac{1}{16 \pi^2}
\left[Y^\ast_{\nu, i \kmax} Y^{T}_{\nu, \kmax j}
m_{\tilde{N},\kmax \kmax}^2 \right. \nnn
&+& \left.
A^\ast_{\nu,i\kmax} A^T_{\nu,\kmax j}+
Y^\ast_{\nu,i\kmax} Y^T_{\nu,\kmax j} m^2_{H_2}\right.\nnn
&+&\left.
Y^\ast_{\nu,i\kmax} [\sum_{k'=1}^{\kmax-1} Y^T_{\nu ,k'j}
m^2_{{\tilde N},\kmax k'}]
+ [\sum_{k=1}^{\kmax-1} Y^\ast_{\nu,ik} m^2_{{\tilde N},k \kmax}
 ] Y^T_{\nu, \kmax j} \right]\nnn
&+&\frac{1}{16 \pi^2}\left[ 2 Y^\ast_{\nu,i\kmax} Y^T_{\nu, \kmax j}
m^2_{{\tilde N},\kmax \kmax} \right. \nnn
&+& \left. (Y^\ast_{\nu,i \kmax} A^T_{\nu, \kmax j}
 {B^{2 \ast}_{\kmax \kmax}}
+ A^\ast_{\nu,i \kmax} Y^T_{\nu, \kmax j} {B^2_{\kmax \kmax}})
\frac{1}{M_{R,\kmax}} \right],
\label{eq:sleptonsoft}
\end{eqnarray}
where we use ${B^2_{ij}}= B_0 M_{R,i} \delta_{ij}$ and both scheme dependent and
independent threshold corrections are included. If $B_0$ is
larger than the other
soft breaking parameters, i.e., $B_0 > A_\nu, m_{\tilde N}$,
the scheme dependent
terms in second bracket in Eq.(\ref{eq:sleptonsoft}) are smaller than the scheme independent terms
in third (final) bracket in Eq.(\ref{eq:sleptonsoft}).
Therefore we can neglect the scheme dependent terms for
the numerical analysis.
We notice that the scheme independent terms are the same as those obtained
from RG analysis given by Eq.(\ref{eq:thresha}).

Similarly, one can obtain the threshold corrections for bilinear part of the Higgs field $H_2$.
Calculating the diagrams (d)(e)(f)(i) in Figs. 2,3, we get one-loop corrections to $m_{H_2}^2$ which
are divided by two contributions as follows.
The scalar loop contributions are given by
\begin{eqnarray}
-i\Sigma^s_{{H_2}} &=&
\frac{1}{2}
\left(A_{\nu,ik}^* A_{\nu,ki}^T + Y_{\nu,ik}^* Y_{\nu,ki}^T M_{R,k}^2 \right)\left[I_2(m_{\tilde{L},ii}^2, M_{h,k}^2) +
I_2(m_{\tilde{L},ii}^2, M_{l,k}^2)\right] \nonumber \\
&+& \frac{1}{2}(Y_{\nu,ik}^* A^T_{\nu,ki}e^{-i\Phi_k} + A_{\nu,ik}^* Y_{\nu,ki}^T e^{i\Phi_k}) M_{R,k}\left[I_2(m_{\tilde{L},ii}^2, M_{h,k}^2)-
I_2(m_{\tilde{L},ii}^2, M_{l,k}^2)\right] \nonumber \\
&+& \frac{1}{2}Y_{\nu,ik}^* Y_{\nu,ki}^T[
2I_1(m^2_{{\tilde L},ii}) + I_1(M_{h,k}^2) + I_1(M_{l,k}^2)] \nonumber \\
&+& \sum_{i \ne j} Y_{\nu,ik}^* Y_{\nu,kj}^T m_{\tilde{L},ji}^2 I_2(m_{\tilde{L},ii}^2, m_{\tilde{L},jj}^2) \nonumber \\
&+& \frac{1}{2} Y_{\nu,ik}^* Y_{\nu,k'i}^T e^{\frac{i}{2}(\Phi_{k}-\Phi_{k'})}\Bigl[m_{h,kk'}^2 I_2(M_{h,k}^2, M_{h,k'}^2)
+m_{l,kk'}^2 I_2(M_{l,k}^2, M_{l,k'}^2) \nonumber \\
&+&i m_{hl,k'k}^2 I_2(M_{h,k'}^2, M_{l,k}^2) -i m_{hl,kk'}^2 I_2(M_{h,k}^2, M_{l,k'}^2)\Bigr]+ i \frac{p^2}{32 \pi^2} Y^\ast_{\nu,ik}Y^T_{\nu,ki}.
\label{eq:fermis}
\end{eqnarray}
The fermion loop contributions are given by
\begin{eqnarray}
 -i\Sigma^f_{{H_2}} &=&  -2\sum_{k=1}^{3}
Y_{\nu,ik}^* Y_{\nu,ik} I_1(M_{R,k}^2) -i\Sigma_{{H_2}}^p,
\label{eq:fermih}
\end{eqnarray}
where
\begin{eqnarray}
 -i\Sigma_{{H_2}}^p &=&  \sum_{k=1}^{3} \frac{i}{16\pi^2}
 Y_{\nu,ik}^* Y_{\nu,ki}^T p^2\left(\bar{\epsilon}^{-1}+\frac{1}{2}
-\ln\frac{M_{R,k}^2}{Q^2} \right)
+\mathcal{O}(M_{R,k}^{-1}).
\end{eqnarray}
The explicit expression of the total contributions to $m^2_{H_2}$
is shown in  Eq.(\ref{eq:higgs_self}).
Then the effective Lagrangian for Higgs ($H_2$) field is
given as,
\begin{eqnarray}
 \mathcal{L}_{\rm eff}^{k_{\rm max}} &=&
z_{H_2}\partial_\mu {H}_{2}^*
\partial^\mu {H}_2 \nnn
&-& \left({m^{2(\kmax)}_{H_2}}(Q^2) +
\delta {m^{2(\kmax)}_{H_2}}(Q^2) +
{\delta^{(\kmax)}_{SD} m^2_{H_2}} +
{\delta^{(\kmax)}_{SI} m^2_{H_2}}
\right){H_2}^\dagger H_2. \label{eq:effective_higgs}
\end{eqnarray}
where
\begin{eqnarray}
z_{H_2}(Q^2)&=&1 -
\sum_{k=1}^{k_{\rm max}}\frac{1}{16\pi^2}
Y_{\nu,ik}^* Y_{\nu,ki}^T
(\ln\frac{M_{R,k}^2}{Q^2}-1), \\
 \delta m_{H_2}^{2(\kmax)}(Q^2) &=& \frac{1}{16\pi^2} \biggl[
\sum_{k=1}^{\kmax} (A_{\nu,ik}^* A_{\nu,ki}^T +
 Y_{\nu,ik}^* Y_{\nu,kj}^T m_{\tilde{L},ji}^2 +
Y_{\nu,ik}^* Y_{\nu,ki}^T
m_{\tilde{N},kk}^2)\ln\frac{M_{R,k}^2}{Q^2} \nnn
&+& \sum_{k\neq k',1}^{\kmax} Y_{\nu,ik}^* Y_{\nu,k'i}^T
m_{\tilde{N},kk'}^2 \ln \frac{{\rm max}(M_{R,k}^2, M_{R,k'}^2)}{Q^2} \biggr] , \\
{\delta^{(\kmax)}_{SD} m^2_{H_2}} &=&
-\frac{1}{16\pi^2} \sum_{k=1}^{\kmax}
(A_{\nu,ik}^* A_{\nu,ki}^T + Y_{\nu,ik}^*
Y_{\nu,kj}^T m_{\tilde{L},ji}^2+
\sum_{k'=1}^{\kmax} Y_{\nu,ik}^* Y_{\nu,k'i}^T
 m_{\tilde{N},kk'}^2) , \\
{\delta^{(\kmax)}_{SI} m^2_{H_2}} &=&
\frac{1}{16\pi^2} \sum_{k=1}^{\kmax}
\left[2 Y_{\nu,ik}^* Y_{\nu,ki}^T m_{\tilde{N},kk}^2 +
(Y_{\nu,ik}^* A_{\nu,ki}^T B_{kk}^{2*} + A_{\nu,ik}^* Y_{\nu,ki}^T B_{kk}^2)\frac{1}{M_{R,k}}
\right] \nnn
&-& \sum_{k=1}^{k_{\rm max}} \left[\sum_{i \ne j} \frac{1}{16 \pi^2}
Y^T_{\nu, kj} m^2_{{\tilde L},ji} Y^\ast_{\nu ik}
\frac{m^2_{{\tilde L},jj} \log
\frac{M_{R,k}^2}{m^2_{{\tilde L},jj}}
-m^2_{{\tilde L},ii}
\log \frac{M_{R,k}^2}{m^2_{{\tilde L},ii}} }
{m^2_{{\tilde L},jj}-m^2_{{\tilde L},ii}} \right] .
\end{eqnarray}

Using Eq.(\ref{eq:effective_higgs}), one can obtain the soft mass of up-type Higgs
including scheme dependent and independent threshold corrections as follows,
\begin{eqnarray}
{m^{2(\kmax-1)}_{H_2}}(M_{R,\kmax})&=&
{m^{2(\kmax)}_{H_2}}(M_{R,\kmax}) \left[1-\frac{1}{16 \pi^2}
Y^\ast_{\nu,i \kmax} Y^T_{\nu, \kmax i} \right]
\nnn
&+&
{\delta^{(\kmax)}_{SD} m^2_{H_2}}-{\delta^{(\kmax-1)}_{SD} m^2_{H_2}}
+
{\delta^{(\kmax)}_{SI} m^2_{H_2}}-
{\delta^{(\kmax-1)}_{SI} m^2_{H_2}} \nnn
&=&{m^{2(\kmax)}_{H_2}}(M_{R,\kmax})\left[1-\frac{1}{16 \pi^2}
Y^\ast_{\nu,i \kmax} Y^T_{\nu, \kmax i}\right] \nnn
&-&\frac{1}{16 \pi^2}
\left[A^\ast_{\nu,i\kmax} A^T_{\nu,\kmax i}+
Y^\ast_{\nu,i \kmax} Y^T_{\nu,\kmax j} m^2_{\tilde{L},ji}
 \right. \nnn
&+& \left. Y^\ast_{\nu,i \kmax} \sum_{k'=1}^{\kmax-1} Y^T_{\nu ,k'i}
m^2_{{\tilde N},\kmax k'}
+ \sum_{k=1}^{\kmax-1} Y^\ast_{\nu,ik} m^2_{{\tilde N},k \kmax}
Y^T_{\nu, \kmax i} \right. \nnn
&+& \left. Y^\ast_{\nu, i\kmax}
Y^{T}_{\nu, \kmax i} m_{\tilde{N},\kmax \kmax}^2 \right]
+\frac{1}{16 \pi^2}\left[
 2 Y^\ast_{\nu,i \kmax} Y^T_{\nu, \kmax i}
m^2_{{\tilde N},\kmax \kmax}  \right. \nnn
&+& \left. (Y^\ast_{\nu,i \kmax} A^T_{\nu, \kmax i}
{B^{2 \ast}_{\kmax \kmax}}
+ A^\ast_{\nu,i \kmax} Y^T_{\nu, \kmax i} {B^2_{\kmax \kmax}})\frac{1}{M_{R,\kmax}} \right]
\nnn
&-& \sum_{i \ne j} \frac{1}{16 \pi^2}
Y^T_{\nu, \kmax j} m^2_{{\tilde L},ji} Y^\ast_{\nu, i \kmax}
\frac{m^2_{{\tilde L},jj} \log
\frac{M_{R,\kmax}^2}{m^2_{{\tilde L},jj}}
-m^2_{{\tilde L},ii} \log \frac{M_{R,\kmax}^2}{m^2_{{\tilde L},ii}} }
{m^2_{{\tilde L},jj}-m^2_{{\tilde L},ii}}.\nnn
\end{eqnarray}
Here, similar to the case of slepton masses, the scheme dependent terms are smaller
than the scheme independent terms.


\section{Lepton Flavor Violating Decays and the relic abundance of  neutralino dark matter}
As we have seen in the previous section, the threshold corrections to slepton and Higgs masses can be so large that they  dominate over
the RGE running effects which are approximately expressed in Eq.(\ref{eq:run1}) and (\ref{eq:run2}).
In this paper, for our numerical calculation, we consider so called minimal supergravity scenario(mSUGRA) where gaugino masses, soft scalar masses and scalar trilinear couplings are universal at the GUT scale.
In particular, we investigate the relic abundance of neutralino dark matter in the focus point region\cite{chan1998,Feng2000,Feng2000a}, which is
one of the regions where the relic density is consistent with WMAP observation.
In the region, the lightest supersymmetric
particle (LSP) is a mixture state of bino and higgsino and the annihilation cross section  for LSP is enhanced due to the appropriate portion of the
higgsino component.
 In the seesaw model without threshold corrections, the focus point region is significantly affected by the neutrino
 Yukawa sector and shift to the energy scale far from electroweak scale when the right-handed Majorana masses are sufficiently large
\cite{Barger2008, Calibbi2007, Kadota2009}.
In SUSY type-I seesaw model based on mSUGRA, the radiative corrections drive the up-type Higgs mass squared to be more negative
than that in MSSM if we do not include the threshold corrections. This feature is due to the presence of RH neutrino and sneutrino sectors.
In this case, the higgsino mass $|\mu|$ is larger than that in MSSM, too.
This in turn leads to larger relic abundance of neutralino dark matter compared to that in MSSM for fixed values of the soft scalar masses
 due to the small portion of the higgsino components in the LSP.
However, if we include large threshold corrections mentioned above, the up-type Higgs mass squared is driven to be less
negative which leads to smaller $|\mu|$.
Therefore the portion of higgsino in the lightest neutralino state becomes large, and
the right amount of relic abundance of neutralino dark matter consistent with WMAP observation can be obtained
even in the parameter space of MSSM excluded by WMAP data \cite{Kang:2009pj}.

Since the threshold corrections mentioned above produce flavor off-diagonal slepton masses, they are new additional source of
lepton flavor violating phenomena.
Since large threshold corrections lead to large flavor off-diagonal slepton masses, we anticipate that the amplitude of
lepton flavor violating processes such as $\tau\rightarrow \mu \gamma$,
$\tau \rightarrow e \gamma$ and $\mu \rightarrow e \gamma$ can be enhanced due to the new source of lepton flavor violation.
Therefore there exists a tension between the branching ratio of LFV decays and the relic abundance of neutralino dark matter.

Higgsino mass parameter $\mu$ is determined by the minimization condition of the Higgs potential  given as,
\begin{eqnarray}
 \frac{1}{2} m_Z^2 = - |\mu|^2 + \frac{m_{H_1}^2(m_Z^2)-m_{H_2}^2(m_Z^2)\tan^2\beta}{\tan^2\beta -1} .
\end{eqnarray}
In the limit of large $\tan\beta$, this condition can be written as
\begin{eqnarray}
 \frac{1}{2} m_Z^2 \approx - |\mu|^2 - m_{H_2}^2(m_Z^2) .
\end{eqnarray}
The radiative corrections to $m_{H_2}^2$ are given as,
\begin{eqnarray}
 \delta m_{H_2}^2 \approx \sum_k \frac{Y_{\nu,ik}^* Y_{\nu,ki}^T}{8\pi^2} \left[
m_0^2 + A_0 B_N^0 - \left(A_0^2 +3 m_0^2\right) \ln \frac{M_{GUT}}{M_{R,k}}
\right], \label{corr2}
\end{eqnarray}
where the first two terms of RH side are the threshold corrections and the last term is the RG running effect.
We see from Eq.(\ref{corr2}) that in the case of the large threshold corrections to $m_{H_2}^2$, the first two contributions
to $ \delta m_{H_2}^2$ dominate over the last one and thus Higgsino mass parameter $|\mu|^2$ becomes small.
This lowers the abundance of neutralino dark matter compared to that of MSSM.
From the constraint from WMAP observation, we can obtain the allowed parameter space.


Now, let us consider the radiative LFV decays and investigate how they are related with the relic abundance of neutralino dark matter.
Branching ratios of the radiative LFV decays are approximately given by \cite{Casas2001}
\begin{eqnarray}
 BR(l_i \rightarrow l_j \gamma) \sim \frac{\alpha^3}{G_F^2} \frac{|m^2_{\tilde{L},ij}|^2}{m_{s}^8}\tan^2\beta, \label{eq:lfv}
\end{eqnarray}
where $m_{s}$ is average SUSY scalar mass. For numerical calculations, we have used the complete formulas which are given by Eq.(51) in
Ref.\cite{Hisano1996}.
As we can see from Eq. (\ref{eq:lfv}), the branching ratios of the LFV decays are proportional to $|m_{\tilde{L},ij}^2|^2$.
Including the threshold corrections,  $m_{\tilde{L},ij}^2$ can be approximately written as
\begin{eqnarray}
 m_{\tilde{L},ij}^2 \approx \sum_k \frac{Y_{\nu,ik}^* Y_{\nu,kj}^T}{8\pi^2} \left[ m_0^2 + A_0 B_N^0
- (A_0^2 +3 m_0^2) \ln \frac{M_{GUT}}{M_{R,k}}\right] .
\label{eq:soft_slepton}
\end{eqnarray}
Here, the first two terms of RH side are the threshold corrections given in Eq.(\ref{eq:th_mlsq}). The last term
 comes from the RG running effects which is obtained by integrating out RGEs for $m_{\tilde{L},ij}^2$. Because the relative sign
 is opposite, both contributions cancel each other when their contributions are comparable. As the $B_N^0$ becomes large, the first two terms
 gets dominant over the last term. It is worthwhile to notice that the size of $B_N^0$ is limited by the constraints from the experiments.




\section{Numerical Calculation}
In this section, we present our numerical calculations of the relic abundance of the neutralino dark matter and the branching ratio of the LFV decays.
For our numerical calculations, we use the micrOMEGAs package\cite{Belanger:2006is,Belanger:2008sj} for the relic abundance and the
SuSpect\cite{Djouadi2007} for RGE running of soft SUSY breaking parameters with appropriate modification. Then we compare our numerical results with the experimental ones.
We use the following upper bound for the branching ratio of the LFV decays : ${\rm BR}(\tau \rightarrow \mu \gamma) <4.4\times 10^{-8}$ \cite{babar2009},  ${\rm BR}(\tau \rightarrow e \gamma)<3.3\times 10^{-8}$ \cite{babar2009}
 and  ${\rm BR}(\mu \rightarrow e \gamma)<1.2\times 10^{-11}$ \cite{mega1999}.

The WMAP observation leads to the relic abundance of the cold dark matter, $\Omega_{CDM} h^2 $ given by\cite{G.2009},
\begin{eqnarray*}
 \Omega_{CDM} h^2 = 0.111^{+0.011}_{-0.015}\ \  (2\sigma) .
\end{eqnarray*}

For numerical calculations, we need to parameterize the neutrino Dirac Yukawa couplings in terms of light neutrino masses and mixing.
From the superpotential, the Lagrangian for the neutrino sector is given as,
\begin{eqnarray}
{\cal L}=-\frac{1}{2} Y^T_{\nu,ki} \overline{N_R^k} l_i \cdot H_2
-\frac{1}{2} Y_{\nu,ik} l^c_i \cdot H_2 {N_R^k}^c
-\frac{1}{2} \overline{N_R^k} M_{R,k} {N_R^k}^c + h.c. .
\end{eqnarray}
After integrating out the heavy Majorana neutrinos, one obtains the
dimension five operator.
\begin{eqnarray}
{\cal L}_{\rm eff}=+\frac{1}{2}
 (\overline{l_{j}^c} \cdot H_2) Y_{\nu,jk}
\frac{1}{M_{R,k}}  Y^T_{\nu, ki} (l_i \cdot  H_2),
\end{eqnarray}
The mass terms for the left-handed light neutrinos are given by
\begin{eqnarray}
 \mathcal{L} = - \frac{1}{2} \overline{\nu_L^c} M_\nu \nu_L + h.c. \ ,
\end{eqnarray}
where the left-handed light neutrinos $\nu$ are presented in the flavor eigenstate, $\nu^T = (\nu_e, \nu_\mu, \nu_\tau)$, and
$M_\nu$ denotes the neutrino mass matrix written as
\begin{eqnarray}
 (M_{\nu})_{ij} = -Y_{\nu,ik} M_{R,k}^{-1}Y_{\nu,kj}^T \left<H_2^0\right>^2 .
\end{eqnarray}
The neutrino mass matrix $M_{\nu}$ can be diagonalized by the unitary matrix $U_{\nu}$ as follows:
\begin{eqnarray}
 U_{\nu}^T M_{\nu} U_{\nu} = {\rm diag}(m_{\nu 1}, m_{\nu 2}, m_{\nu 3}) \equiv M_{\nu D}.
\end{eqnarray}
The unitary matrix $U_\nu$ can be identified to $U_{MNS}$, defined by
\begin{eqnarray}
U_{MNS} &=& \left(
\begin{array}{ccc}
c_{12}c_{13} & s_{12}c_{13} & s_{13}e^{-i\delta} \\
-s_{12}c_{23}-c_{12}s_{23}s_{13}e^{i\delta} & c_{12}c_{23}-s_{12}s_{23}s_{13}e^{i\delta} & s_{23}c_{13} \\
s_{12}s_{23}-c_{12}c_{23}s_{13}e^{i\delta} & -c_{12}s_{23}-s_{12}c_{23}s_{13}e^{i\delta} & c_{23}c_{13}
\end{array}
\right) \nonumber \\
&& \times {\rm diag}\left(e^{i\alpha_1/2}, e^{i\alpha_2/2}, 1\right) .
\end{eqnarray}
where $\delta, \alpha_1, \alpha_2$ are CP-violating phases, $c_{ij}=\cos(\theta_{ij})$ and $s_{ij}=\sin(\theta_{ij})$ with mixing angles $\theta_{ij}$.
The current experimental values and bounds for the mixing angles are
\begin{eqnarray}
 \sin^2(2\theta_{12}) = 0.87\pm0.03, \  \sin^2(2\theta_{23}) > 0.92, \  \sin^2(2\theta_{13}) < 0.19 .
\end{eqnarray}
Since we do not consider the CP-violation, the phases are set to zero.
In our numerical calculations, we take $ \theta_{12}=0.6$, $\sin\theta_{23}=\cos\theta_{23}=1/\sqrt{2}$ and $\sin\theta_{13}=0$.
It is well known that the neutrino Dirac Yukawa matrix can be written in terms of $U_{MNS}$ and the diagonal forms of the mass matrices as follows
 \cite{Casas2001};
\begin{eqnarray}
 Y_{\nu}^T =(i) \frac{1}{\left<H_2^0\right>} \sqrt{M_R} R \sqrt{M_{\nu D}} U_{MNS}^\dag ,
\label{eq:Y}
\end{eqnarray}
where R is a complex orthogonal matrix. We assume that $R=1$ for the sake of simplicity in our numerical calculations.

The combination $(Y_{\nu}^* Y_{\nu}^T)_{ij}$ can be written as
\begin{eqnarray}
 (Y_{\nu}^* Y_{\nu}^T)_{ij} = \frac{1}{\left<H_2^0\right>^2} U_{\nu,ik} m_{\nu k} M_{R,k} U_{\nu,kj}^\dag .
\end{eqnarray}
Assuming that neutrino masses are subject to normal hierarchy, we can take
\begin{eqnarray}
 (m_{\nu1}, m_{\nu2}, m_{\nu3}) = (0, \sqrt{\Delta m_{21}^2}, \sqrt{\Delta m_{21}^2+\Delta m_{32}^2}) ,
\end{eqnarray}
where $\Delta m_{21}^2$ and $\Delta m_{32}^2$ are mass difference of neutrinos, and their experimental results are
$\Delta m_{21}^2=(7.57\pm0.20)\times 10^{-5}{\rm eV}^2$ and $\Delta m_{32}^2=(2.43\pm0.13)\times 10^{-3}{\rm eV}^2$.
We also assume that $M_{R,1} \ll M_{R,2} \ll M_{R,3}$. In this case,
\begin{eqnarray}
 (Y_{\nu}^* Y_{\nu}^T)_{ij} \sim \frac{1}{\left<H_2^0\right>^2} U_{\nu,i3} m_{\nu 3} M_{R,3} U_{\nu,3j}^\dag .
\end{eqnarray}
Thanks to $\sin\theta_{13} = 0$, $(Y_{\nu}^* Y_{\nu}^T)_{13} \ll (Y_{\nu}^* Y_{\nu}^T)_{23}$. As we can see from eq.(\ref{eq:soft_slepton}),
this implies that $m_{\tilde{L},13}^2 \ll m_{\tilde{L},23}^2$.
Therefore, we anticipate that ${\rm BR}(\tau\rightarrow e \gamma) \ll
 {\rm BR}(\tau\rightarrow \mu \gamma)$ despite the current experimental bounds are of same order.
So, we do not present  ${\rm BR}(\tau\rightarrow e\gamma)$ in this work.

\begin{figure}[htbp]
\includegraphics[width=15cm]{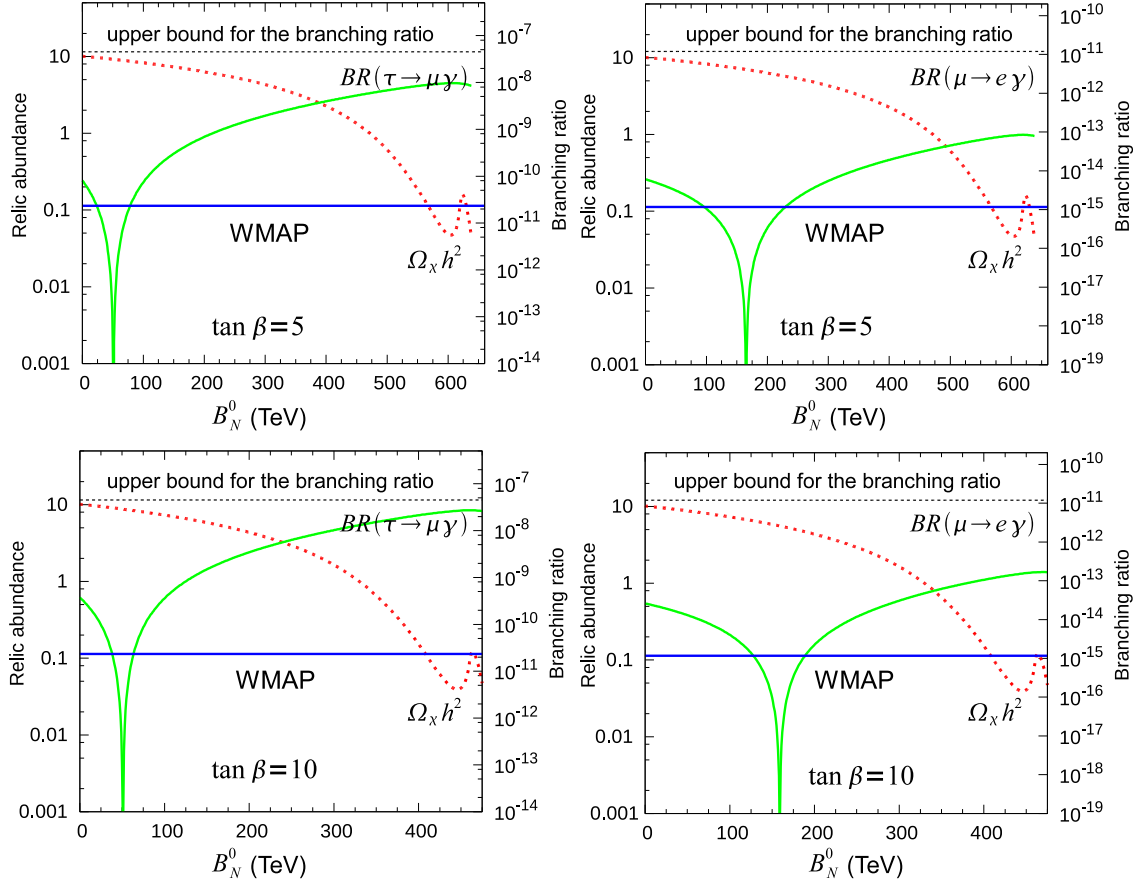}
\caption{The branching ratios of LFV decays and the relic abundance of the lightest neutralino are shown as functions of $B_N^0$ with hierarchical neutrino
mass case.
Two left panels show  $BR(\tau\rightarrow\mu\gamma)$ {\it vs.} the relic abundance and two right panels show $BR(\mu\rightarrow e\gamma)$
{\it vs.} the relic abundance. In each panel, green solid and red dotted curves represent the corresponding branching ratio
and the relic abundance of the lightest neutralino, respectively.
The gray dotted (upper) and blue solid (lower) straight lines represent the upper bound on the branching ratio given by experiments
and the abundance of the dark matter obtained from WMAP, respectively.
We take
$(M_{R,1}, M_{R,2}, M_{R,3})=(10^{10}, 10^{12}, 10^{14})
 {\rm GeV}$ and $\tan\beta=5$ for upper two panels and $\tan\beta=10$ for lower two panels.
We also take $m_0=1{\rm TeV},
A_0=300{\rm GeV}, m_{1/2}=300 {\rm GeV}$ and $\mu>0$.
}
\label{fig:lfv_vs_relic}
\end{figure}

\begin{figure}[htbp]
\includegraphics[width=15cm]{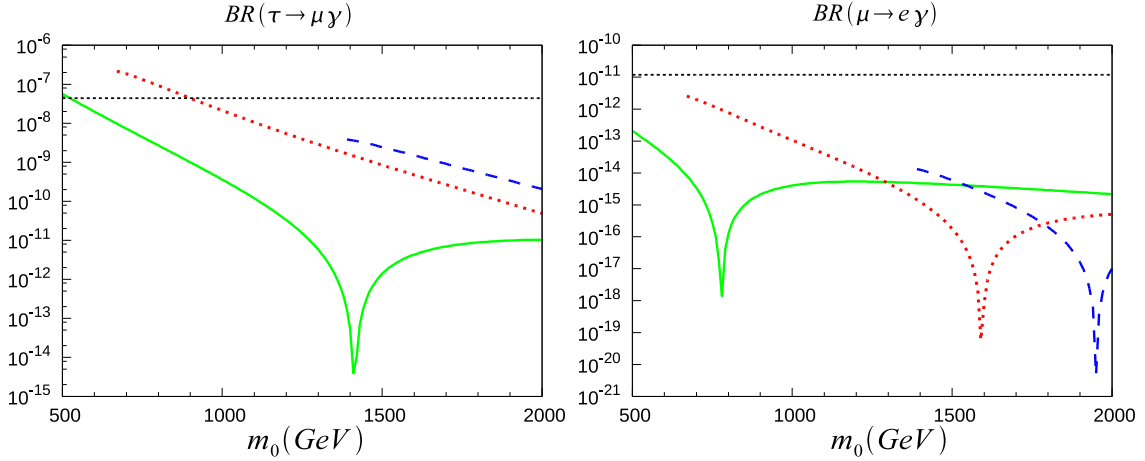}
\caption{The branching ratios of LFV decays
are shown as functions
of universal soft scalar mass $m_0$ for hierarchical neutrino mass case.
The left panel shows $BR(\tau \to \mu \gamma)$
and the right panel shows $BR(\mu\rightarrow e \gamma$).
The gray dotted lines correspond to
the experimental upper bound on the branching fractions.
Green solid, red dotted and blue broken lines are calculated with
$B_N^0=(100,400,600) {\rm TeV}$,respectively.
We choose $m_{1/2}=300 {\rm GeV},
A_0 = 300{\rm GeV}, \tan\beta=10$.}
\label{fig:lfv_changes.eps}
\end{figure}

Fig.\ref{fig:lfv_vs_relic} shows how both the relic relic abundance and the branching ratios of the LFV decays
 simultaneously depend on the parameter $B_N^0$ when light neutrino masses are hierarchical.
We take universal soft scalar mass
 $m_0$ to be $1{\rm TeV}$. Here, note that lightest SUSY particle is the lightest neutralino.
 We also take $\tan\beta=5(10)$ in the upper (lower) panels.
The other input values of the parameters
we take are presented in the caption of Fig.\ref{fig:lfv_vs_relic}.
The left(right) two panels in Fig.\ref{fig:lfv_vs_relic}
$BR(\tau\rightarrow\mu\gamma)$ ($BR(\mu\rightarrow e\gamma)$) {\it vs.}
the relic abundance
 as a function of $B^0_N$ is shown.
In each panels, green solid and red dotted curves present
the branching ratio of LFV
and the relic abundance of the lightest neutralino, respectively.
The gray dotted (upper) and blue solid (lower) horizontal lines
show the current upper bound on the branching ratio from experiments
and the relic abundance of the dark matter obtained from WMAP, respectively.
When $B_N^0$ is so large that the prediction of $\Omega_{\chi}h^2$ fit to the observed abundance of the dark matter,
$BR(\tau\rightarrow \mu \gamma)$ and $BR(\mu\rightarrow e \gamma)$ are predicted to be quite large.
In this case, the prediction of $BR(\tau\rightarrow \mu \gamma)$ almost reaches to the experimental bound.
Although there will be a chance to probe the LFV decays in the case of large value of $B_N^0$ in future experiments,
the size of $B_N^0$ is limited by constraint coming from the relic abundance of neutralino dark matter.
In Fig.\ref{fig:lfv_changes.eps},
we also show how the branching ratios
of LFV depend on the universal soft scalar mass $m_0$.
%
Here, there exist the valleys where the branching ratios are extremely suppressed, which are occurred due to
cancellation between RG running effects and the threshold corrections.
This is expected from
Eq.(\ref{eq:soft_slepton}), since the sign of the contribution from RG
running effect to the soft SUSY breaking terms for slepton is opposite to
the threshold correction. Therefore as $m_0$ increases, the large threshold
correction from $B_N^0$ is cancelled and branching ratios for LFV become
small.
As can be seen from Fig. \ref{fig:lfv_changes.eps}, the dotted (dashed) line starts from the point corresponding to $m_0=700 (1400)$ GeV
below which electroweak symmetry is not radiatively broken due to too large corrections to $m_{H_2}^2$.
In the case of large $B_N^0$ such as $400 \sim 600$ TeV,
 the branching ratio for $\tau  \rightarrow \mu \gamma$ and $\mu  \rightarrow e \gamma$
become so sizable that the LFV decay could be detected even when $m_0$ is larger than $1 {\rm TeV}$.

In the case of inverted hierarchy, three neutrino masses are given as
\begin{eqnarray}
 (m_{\nu3}, m_{\nu1}, m_{\nu2}) = (m_{\nu3}, \sqrt{m_{\nu3}^2 +\Delta m_{32}^2-\Delta m_{21}^2}, \sqrt{m_{\nu3}^2 +\Delta m_{32}^2}),
\end{eqnarray}
and $m_{\nu3} < m_{\nu1} < m_{\nu2}$.
The threshold corrections to $\delta m_{H_2}^2$ is proportional to ${\rm Tr}(Y_{\nu}^* Y_{\nu}^T)$.
${\rm Tr}(Y_{\nu}^* Y_{\nu}^T)$ is approximately given as
\begin{eqnarray}
 {\rm Tr}(Y_{\nu}^* Y_{\nu}^T) &\approx& \Bigl[(|U_{\nu,12}|^2 + |U_{\nu,22}|^2 + |U_{\nu,32}|^2) m_{\nu2} M_{R,2} \nonumber \\
&&+ (|U_{\nu,23}|^2 + |U_{\nu,33}|^2) m_{\nu3} M_{R,3} \Bigr]/\left<H_2^0\right>^2. \label{YY}
\end{eqnarray}
In the case of ${m_{\nu3}}/{m_{\nu2}} \ll M_{R,2}/M_{R,3}$, the first term in Eq.(\ref{YY}) is dominant.
The ratio of ${\rm Tr}(Y_{\nu}^* Y_{\nu}^T)$ in the inverted hierarchy to
that in the normal hierarchy is
\begin{eqnarray}
 {{\rm Tr}(Y_{\nu}^* Y_{\nu}^T)}/{{\rm Tr}(Y_{\nu}^* Y_{\nu}^T)^{NOR}} \sim {M_{R,2}}/{M_{R,3}} .
\end{eqnarray}
Therefore threshold corrections to $m_{H_2}^2$ are much smaller than in the normal hierarchy case.
This leads to rather larger abundance of the neutralino dark matter.
On the other hand, the term $\left(Y_{\nu}^* Y_{\nu}^T\right)_{21}$ is written as
\begin{eqnarray}
 \left(Y_{\nu}^* Y_{\nu}^T\right)_{21} \approx U_{\nu,22}
U^\dagger_{\nu,21} m_{\nu,2} M_{R,2}/\left<H_2^0\right>^2.
\end{eqnarray}
This term is obviously larger than that in the normal hierarchy. In particular, for $m_{\nu3} \sim 0$,
\begin{eqnarray}
 {\left(Y_{\nu}^* Y_{\nu}^T\right)_{21}}/{\left(Y_{\nu}^* Y_{\nu}^T\right)^{NOR}_{21}} \sim 10.
\end{eqnarray}
Therefore we naively expect that $BR(\mu\rightarrow e\gamma)$ becomes larger by 2 order of magnitude than that in the normal hierarchy
for the same value of $B_N^0$. As a result, it becomes difficult to satisfy the constraint from the current bounds on $BR(\mu\rightarrow e\gamma)$ and the relic abundance simultaneously.

\begin{figure}[htbp]
\includegraphics[width=15cm]{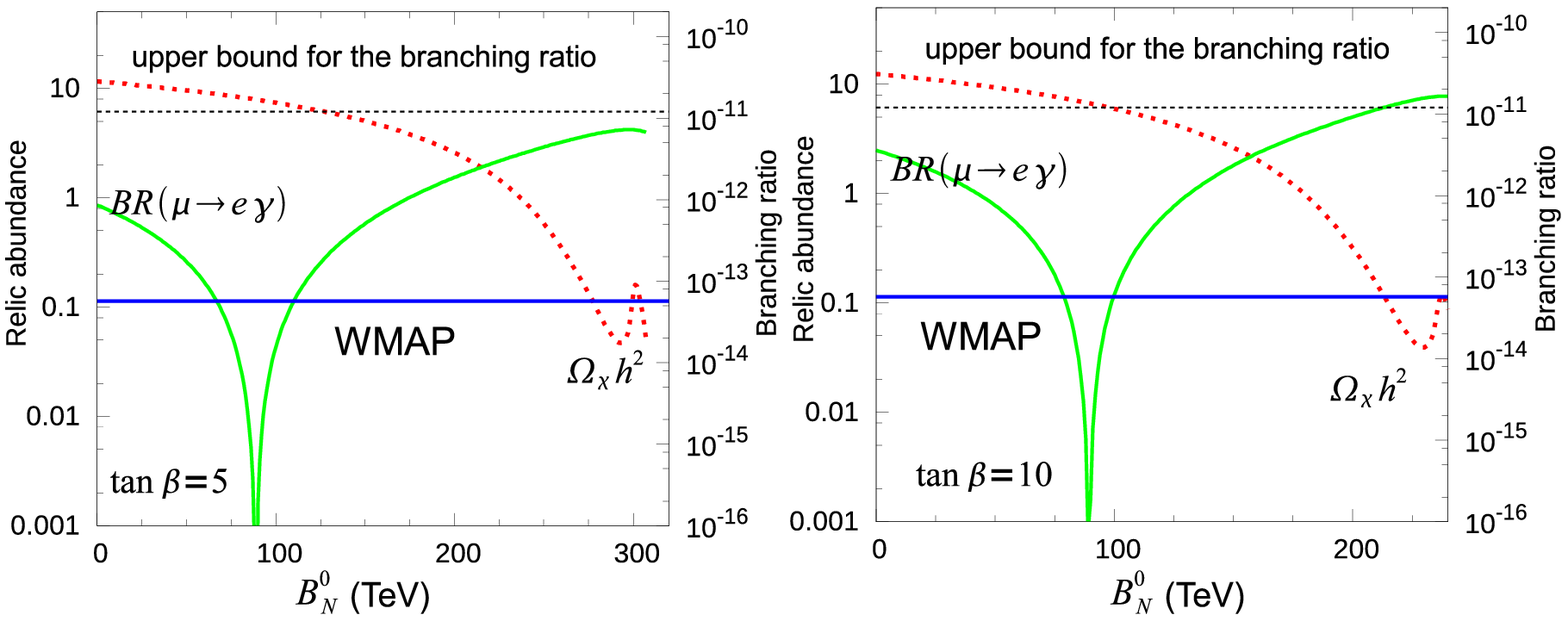}
\caption{Br($ \mu \to e \gamma$) vs. the relic abundance
as function of $B_N^0$ for degenerate neutrino case.
The green solid and
red dotted curves represent the branching ratio
and the relic abundance of the lightest neutralino, respectively.
The gray dotted (upper) and blue solid (lower) straight lines represent the upper bound on the branching ratio given by experiment.
The parameters are the
same as those in Fig.\ref{fig:lfv_vs_relic}.
}
\label{fig:lfv_vs_relic2}
\end{figure}

When the neutrino masses are almost degenerate, i.e. $m_{\nu1} \sim m_{\nu2} \sim m_{\nu3}$, $(Y_{\nu}^*Y_{\nu}^T)_{21}$ is larger than that in the normal hierarchy, which makes the prediction of $BR(\mu\rightarrow e\gamma)$ in this case substantially enhanced compared to the normal hierarchical
case, and thus the constraint of $BR(\mu\rightarrow e\gamma)$
becomes more severe when we consider the constraint of the relic abundance simultaneously.
We present the branching ratios of the LFV decay $\mu\rightarrow e \gamma$ and the relic abundance of the lightest
neutralino as functions of $B_N^0$ in Fig.\ref{fig:lfv_vs_relic2}.
The different figures correspond to different values of $\tan \beta$. As $\tan\beta$ increases, larger value of $B_N^0$
is preferred to accommodate both $BR(\mu\rightarrow e \gamma)$ and  the relic abundance of dark matter candidate.


\section{Conclusion}
We have investigated the LFV radiative decays and the relic abundance of the neutralino dark matter in the SUSY seesaw model.
We have carefully derived the threshold corrections to the flavor off-diagonal
elements of slepton mass matrix and up-type Higgs mass squared  and found that
they can be so large in the case of large $B_N^0$ that the branching ratios of the LFV decays and the relic abundance of neutralino dark matter
can be significantly affected.
Our numerical results show that there are parameter regions where the prediction of the relic abundance of neutralino dark matter
is consistent with WMAP observation and the branching ratios of LFV radiative decays can be enhanced so as for them to be probed in
future experiments \cite{Aushev:2010bq,Bona:2007qt}.
Although the origin of such large $B_N^0$ is unclear, if such large B-term exists, the branching ratios of LFV decays
 are significantly enhanced even when $m_0$ is not small.
 Therefore, the masses of scalar supersymmetric particles are not necessarily small so that the branching ratios of LFV decays
can be testable in future experiment, which is distinctive feature of this scenario.

\section{ACKNOWLEDGEMENT}
The work of T. M. is supported by KAKENHI, Grant-in-Aid for Scientific Research
on Priority Areas, New Development of Flavor Physics No.20039008 from
MEXT and Grand-in-Aid for Scientific
Research(C) No.22540283 from JSPS, Japan. N.Y. is supported by Grand-in-Aid for Scientific Research,
 No.22-7585 from JSPS, Japan.
The work of S.K.K. is supported in part by Basic Science Research Program through the NRF of Korea
funded by MOEST (2009-0090848).

\appendix
\section{Renormalization Group Equation}
In this appendix, we present one-loop renormalization group equations for $SU(2)$ slepton masses and Higgs mass squared parameters including
the threshold effects. The one-loop RGE for the parameter $B^2_{ij}$ is also presented.
Here, we have omitted the same contributions as in MSSM.
The RGEs for $SU(2)$ slepton masses are given by
\begin{eqnarray}
 (16\pi^2) Q \frac{d m_{\tilde{L},ij}^2}{d Q} &=&
\sum_{k}
\left[ Y_{\nu,ik}^* m_{\tilde{N},kk}^2 Y_{\nu,kj}^T +
\sum_{k'=1}^{k-1} (Y_{\nu,ik'}^* m_{\tilde{N},k'k}^2 Y_{\nu,kj}^T
+ Y_{\nu,ik}^* m_{\tilde{N},kk'}^2 Y_{\nu,k'j}^T )\right] \nnn
&& \times \left[\theta(Q^2-M_{h,k}^2)+\theta(Q^2-M_{l,k}^2)\right] \nnn
&+& \sum_k \left[A_{\nu,ik}^* A_{\nu,kj}^T + m_{22}^2 Y_{\nu,ik}^* Y_{\nu,kj}^T \right]
\left[\theta(Q^2-M_{h,k}^2)+\theta(Q^2-M_{l,k}^2)\right] \nnn
&-& 2 |\mu|^2 \sum_k Y_{\nu,ik}^* Y_{\nu,kj}^T \theta(Q^2-M_{R,k}^2) \nnn
&+& \sum_{k,k'} \left[Y_{\nu,ik}^* Y_{\nu,kk'}^T m_{\tilde{L},k'j}^2
+ m_{\tilde{L},ik'}^2 Y_{\nu,k'k}^* Y_{\nu,kj}^T \right] \theta(Q^2-M_{R,k}^2) \nnn
&+& 2 \sum_{k} Y_{\nu,ik}^* Y_{\nu,kj}^T M_{R,k}^2
\left[ \theta(Q^2-M_{h,k}^2)+\theta(Q^2-M_{l,k}^2)-2\theta(Q^2-M_{R,k}^2)\right] \nnn
&+& \sum_k \left[Y_{\nu,ik}^* A_{\nu,kj}^T e^{-i\Phi_k} + A_{\nu,ik}^* Y_{\nu,kj}^T e^{i\Phi_k}\right] M_{R,k} \nnn
&& \times \left[\theta(Q^2-M_{h,k}^2) - \theta(Q^2-M_{l,k}^2)\right] \nnn
&+& \sum_{k}\left[ Y_{\nu,ik}^* |B_{N,k}| M_{R,k} Y_{\nu,kj}^T
+ \sum_{k'=1}^{k-1} \left(Y_{\nu,ik'}^* B_{k'k}^2 e^{-i\Phi_k} Y_{\nu,kj}^T
+ Y_{\nu,ik}^* B_{kk'}^{2*} e^{i\Phi_k} Y_{\nu,k'j}^T \right)
\right] \nnn
&& \times \left[\theta(Q^2-M_{h,k}^2)-\theta(Q^2-M_{l,k}^2)\right],
\end{eqnarray}
where $m_{22}^2=m_{H_2}^2+|\mu|^2$.
The RGEs for Higgs mass squared parameters are given by
\begin{eqnarray}
 (16\pi^2)Q\frac{d m_{H_2}^2}{d Q} &=& \sum_{i,k}
\left[ Y_{\nu,ik}^* m_{\tilde{N},kk}^2 Y_{\nu,ki}^T +
\sum_{j=1}^{k-1} (Y_{\nu,ij}^* m_{\tilde{N},jk}^2 Y_{\nu,ki}^T
+ Y_{\nu,ik}^* m_{\tilde{N},kj}^2 Y_{\nu,ji}^T )\right] \nnn
&& \times \left[\theta(Q^2-M_{h,k}^2)+\theta(Q^2-M_{l,k}^2)\right] \nnn
&+& \sum_{i,j,k} (Y_{\nu,ki}^T m_{\tilde{L},ij}^2 Y_{\nu,jk}^*) \left[\theta(Q^2-M_{h,k}^2)+\theta(Q^2-M_{l,k}^2)\right] \nnn
&+& \sum_{i,k} (A_{\nu,ik}^* A_{\nu,ki}^T) \left[\theta(Q^2-M_{h,k}^2)+\theta(Q^2-M_{l,k}^2)\right] \nnn
&+& 2m_{H_2}^2 \sum_{i,k} Y_{\nu,ik}^* Y_{\nu,ki}^T \, \theta(Q^2-M_{R,k}^2) \nnn
&+& 2 \sum_{i,k} Y_{\nu,ik}^* Y_{\nu,ki}^T M_{R,k}^2 \left[\theta(Q^2-M_{h,k}^2)+\theta(Q^2-M_{l,k}^2)
- 2 \theta(Q^2-M_{R,k}^2)\right] \nnn
&+& 2 \sum_{i,k} {\rm Re}(Y_{\nu,ik}^* A_{\nu,ki}^T e^{-i\Phi_k}) M_{R,k}
\left[\theta(Q^2-M_{h,k}^2)-\theta(Q^2-M_{l,k}^2)\right] \nnn
&+& \sum_{i,k}\left[ Y_{\nu,ik}^* |B_{N,k}| M_{R,k} Y_{\nu,ki}^T
+ 2 \sum_{j=1}^{k-1} {\rm Re}\left(Y_{\nu,ij}^* B_{jk}^2 e^{-i\Phi_k} Y_{\nu,ki}^T \right)
\right] \nnn
&& \times \left[\theta(Q^2-M_{h,k}^2)-\theta(Q^2-M_{l,k}^2)\right],
\end{eqnarray}

\begin{eqnarray}
(16\pi^2)Q\frac{d m_{11}^2}{d Q} &=& |\mu|^2
\sum_{i,k}|Y_{\nu,ik}|^2
[\theta(Q^2-M_{h,k}^2)+\theta(Q^2-M_{l,k}^2)], \label{m11}
\end{eqnarray}
where $m_{11}^2 \equiv |\mu|^2+m_{H_1}^2$.
{\bf We note that the RH-side of Eq.(\ref{m11}) corresponds to
the contribution from wave function renormalization.}

\begin{eqnarray}
 (16\pi^2)Q\frac{d m^2_{H_1 H_2}}{d Q} &=&
\sum_{i,k}\mu |Y_{\nu,ik}|^2 e^{i \Phi_k} M_{R,k}
[\theta(Q^2-M_{h,k}^2)-\theta(Q^2-M_{l,k}^2)] \nonumber \\
&+&\sum_{i,k} \mu Y_{\nu, ik}^\ast  A_{\nu,ik}
[\theta(Q^2-M_{h,k}^2)+\theta(Q^2-M_{l,k}^2)] \nonumber \\
&+& \sum_{i,k} |Y_{\nu,ik}|^2 m^2_{H_1 H_2} \theta(Q^2-M_{R,k}^2).
\end{eqnarray}

The RGE for $B^2_{ij}$ is given as,
\begin{eqnarray}
 (16\pi^2) Q \frac{d B^2_{ij}}{d Q} &=& 4 \left[ A_{\nu,ik}^T Y_{\nu,kj}^* M_{R,j} + M_{R,i} Y_{\nu,ik}^\dag A_{\nu,kj}\right] \nonumber \\
&+& 2 \left(Y_\nu^T Y_\nu^\ast \right)_{ik} (B^2)_{kj}
+ 2 (B^2)_{ik} \left(Y_\nu^T Y_\nu^\ast \right)_{jk}.
\label{eq:RGEs_B}
\end{eqnarray}
The RG equation for $M_R$ can be found,for instance,
in
\cite{Masina:2003wt},
\begin{eqnarray}
(16\pi^2) Q \frac{d M_{R,ij}}{d Q} &=&
2 \left(Y_\nu^T Y_\nu^\ast \right)_{ik} M_{R,kj}
+ 2 M_{R,ik} \left(Y_\nu^T Y_\nu^\ast \right)_{jk}.
\label{eq:RGEs_MR}
\end{eqnarray}

\section{Diagram calculation}
By summing the self-energies from both fermion and
scalar loop diagrams, the one loop contribution to
the slepton mass squared is given by,
\begin{eqnarray}
 -i\Sigma_{\tilde{L},ij} &=& -\frac{i}{16\pi^2}\Bigl[
-(\bar{\epsilon}^{-1}+1)(A_{\nu,ik}^* A_{\nu,kj}^T +
Y_{\nu,ik}^* Y_{\nu,kj}^T (m_{H_2}^2+ m_{\tilde{N},kk}^2) \nnn
&+&\sum_{k \ne k'} Y_{\nu,ik}^* Y_{\nu,k'j}^T m_{\tilde{N},kk'}^2) \nnn
&+&  (A_{\nu,ik}^* A_{\nu,kj}^T + Y_{\nu,ik}^* Y_{\nu,kj}^T m_{H_2}^2 + Y_{\nu,ik}^* Y_{\nu,kj}^T m_{\tilde{N},kk}^2)\ln\frac{M_{R,k}^2}{Q^2} \nnn
&+& \sum_{k \ne k'} Y_{\nu,ik}^* Y_{\nu,k'j}^T m_{\tilde{N},kk'}^2 \ln \frac{{\rm max}(M_{R,k}^2, M_{R,k'}^2)}{Q^2} \nnn
&+& 2 Y_{\nu,ik}^* Y_{\nu,kj}^T m_{\tilde{N},kk}^2 + (Y_{\nu,ik}^* A_{\nu,kj}^T B_{kk}^{2*} + A_{\nu,ik}^* Y_{\nu,kj}^T B_{kk}^2)\frac{1}{M_{R,k}} \nnn
&+& \sum_{k<k'} (Y_{\nu,ik}^* Y_{\nu,k'j}^T B_{kk'}^2 B_{k'k'}^{2*}+ Y_{\nu,ik'}^* Y_{\nu,kj}^T B_{k'k}^{2*} B_{k'k'}^2)\frac{1}{M_{R,k'}^2}
\Bigr] \nnn
&+& \frac{i}{16\pi^2} Y_{\nu,ik}^* Y_{\nu,kj}^T p^2
[\bar{\epsilon}^{-1}+1-\ln\frac{M_{R,k}^2}{Q^2}].
\label{eq:slepton_self}
\end{eqnarray}
The contribution to the up type Higgs ($H_2$) mass squared is
given by,
\begin{eqnarray}
-i\Sigma_{H_2} &=& -\frac{i}{16\pi^2}\Bigl[
-(\bar{\epsilon}^{-1}+1)(A_{\nu,ik}^* A_{\nu,ki}^T +
Y_{\nu,ik}^* Y_{\nu,ki}^T (m_{{\tilde L},ii}^2+m_{{\tilde N},kk}^2)\nnn
&+& \sum_{k \ne k'}Y_{\nu,ik}^*
Y_{\nu,k'i}^T m_{\tilde{N},kk'}^2
+ \sum_{i \ne j} Y^T_{\nu,kj}m^2_{{\tilde L},ji} Y^\ast_{\nu, ik}) \nnn
&+&  (A_{\nu,ik}^* A_{\nu,ki}^T +
Y_{\nu,ik}^* Y_{\nu,k j}^T m_{{\tilde L},ji}^2 + Y_{\nu,ik}^*
Y_{\nu,ki}^T m_{\tilde{N},kk}^2)\ln\frac{M_{R,k}^2}{Q^2} \nnn
&+& \sum_{k \ne k'} Y_{\nu,ik}^* Y_{\nu,k'i}^T m_{\tilde{N},kk'}^2
\ln \frac{{\rm max}(M_{R,k}^2, M_{R,k'}^2)}{Q^2} \nnn
&-& \sum_{i \ne j}
Y^T_{\nu, kj} m^2_{{\tilde L},ji} Y^\ast_{\nu ik}
\frac{m^2_{{\tilde L},jj} \log
\frac{M_{R,k}^2}{m^2_{{\tilde L},jj}}
-m^2_{{\tilde L},ii} \log \frac{M_{R,k}^2}{m^2_{{\tilde L},ii}} }
{m^2_{{\tilde L},jj}-m^2_{{\tilde L},ii}} \nnn
&+& 2 Y_{\nu,ik}^* Y_{\nu,ki}^T m_{\tilde{N},kk}^2 +
(Y_{\nu,ik}^* A_{\nu,ki}^T B_{kk}^{2*} + A_{\nu,ik}^* Y_{\nu,ki}^T B_{kk}^2)\frac{1}{M_{R,k}} \nnn
&+& \sum_{k<k'} (Y_{\nu,ik}^* Y_{\nu,k'i}^T B_{kk'}^2 B_{k'k'}^{2*}+ Y_{\nu,ik'}^* Y_{\nu,ki}^T B_{k'k}^{2*} B_{k'k'}^2)\frac{1}{M_{R,k'}^2}
\Bigr] \nnn
&+& \frac{i}{16\pi^2}
 Y_{\nu,ik}^*
Y_{\nu,ki}^T p^2[\bar{\epsilon}^{-1}+1-\ln\frac{M_{R,k}^2}{Q^2}].
\label{eq:higgs_self}
\end{eqnarray}
\newpage 
\section{Approximate solutions of the renormalization
group equations for $M_{R}$ and $B^2$.}
As stated below Eq.(\ref{eq:B}),
although the renormalization group running may induce
the large flavor off-diagonal
contribution to $B^2_{ij}$ $(i\ne j)$ at a lower mass scale,
by switching the basis of Majorana mass matrix $M_R$
to the diagonal basis, we can keep $ B^2_{ij}$ in the basis
 almost diagonal because off-diagonal elements
$B^2_{ij} (i \ne j)$ are doubly suppressed by a factor of
$\frac{A_0}{B_0}$ and one loop suppressed factor.
The effect of
the small off-diagonal elements in $B^2_{ij}$ on slepton
soft breaking term turns out to be smaller than
the leading threshold corrections given in Eq.(\ref{eq:th_mlsq}).

In this appendix, we first show that the large
radiatively generated
off-diagonal elements of $B^2_{ij}$ corresponding to the
second term of Eq.(\ref{eq:B}) are
indeed rotated away in the diagonal basis for $M_{R}$.
To show this, the renormalization group equations for $B^2$
and $M_R$ are solved in perturbative way, i.e., we
use the approximation
so that in the RH side of the renormalization
group equations, all the couplings
$A_\nu$ , $Y_\nu$ and mass $M_R$ are scale
independent constants defined at GUT scale where the
initial conditions for renormalization group equations
are imposed. We also show that off-diagonal elements of
the third term of Eq.(\ref{eq:B}) remain even
after the rotation and are numerically
small compared with the leading diagonal elements.

The solutions for Eq.(\ref{eq:RGEs_B}) and
Eq.(\ref{eq:RGEs_MR})
at $Q=M_{R,3}$ with the boundary
conditions in Eq.(\ref{eq:universal_masses}),
are given as,
\begin{eqnarray}
M_{R ij}(M_{R,3})&=&M_{R,i} \delta_{ij}-2 (H_{ij} M_{R,j}+ H_{ji} M_{R,i})
t^{03},  \\
B^2_{ij}(M_{R,3})&=&
B_0 [M_{R,i} \delta_{ij}-2 (H_{ij} M_{R,j}+ H_{ji} M_{R,i})t^{03}]
- 4 A_0 (H_{ij} M_{R,j}+ H_{ji} M_{R,i}) t^{03} \nonumber \\
&=& (B_0+2 A_0) M_{R}(M_{R,3})_{ij} -2 A_0 M_{R,i} \delta_{ij},
\label{eq:solutions}
\end{eqnarray}
where $t^{03}=\frac{1}{16 \pi^2} \log \frac{M_{GUT}}{M_{R,3}}$
and $H=Y_\nu^T Y_\nu^\ast$.
Since the first term of $B^2$ in Eq.(\ref{eq:solutions})
is proportional to the running mass matrix
of the heavy Majorana neutrinos, it is also diagonal
in the diagonal basis for
$M_{R}$.  As for the second term proportional to $A_0$,
it is changed into the non-diagonal
one.
To derive
the unitary matrix $O$ diagonalizing $M_{R}(Q=M_{R,3})$
approximately, we first write the mass matrix
at $Q=M_{R,3}$ in the matrix form as,
\begin{eqnarray}
&& M_{R}(Q=M_{R,3}) \nonumber \\
&=& \left(\begin{array}{ccc}
M_{R1}(M_{R,3})& -2H_{12}t^{03}(M_{R,1}+M_{R,2})
& -2H_{13} t^{03}(M_{R,1}+M_{R,3}) \\
-2H_{12}t^{03}(M_{R,1}+M_{R,2}) & M_{R2}(M_{R,3})
& -2H_{23}t^{03}(M_{R,2}+M_{R,3})\\
-2H_{13} t^{03}(M_{R,1}+M_{R,3}) & -2H_{23}t^{03}(M_{R,2}+M_{R,3})
& M_{R3}(M_{R3})\end{array} \right), \nonumber \\
\label{eq:MMR3}
\end{eqnarray}
where we have used the property $H_{ij}=H_{ji}$ since CP is assumed to be invariant.
The diagonal elements at $Q=M_{R,3}$
are given by
\begin{eqnarray}
M_{Ri}(Q=M_{R,3})=M_{R,i}(1-4 H_{ii} t^{03}).
\end{eqnarray}
One can find the matrix $O$ with
which the mass matrix Eq.(\ref{eq:MMR3}) is diagonalized as,
\begin{eqnarray}
O(M_{R,3})M_{R}(Q=M_{R,3}) O^T(M_{R,3})=D(Q=M_{R,3}),
\label{eq:diag}
\end{eqnarray}
where $D$ is the diagonal matrix. The rotation given
above corresponds to changing the basis $N^c \to O^T N^c$.
In the new basis, $B^2$ is given by,
\begin{eqnarray}
B^2_{{\rm new} ij} \equiv
(O B^2 O^T)_{ij}&=& \{(B_0+2 A_0) D_i(Q=M_{R,3})-2 A_0 M_{R,3} \}
\delta_{ij}\nonumber \\
&-& 2 A_0
\sum_{k=1}^2
(O_{ik}(M_{R,k}-M_{R,3})O_{jk}).
\label{eq:Bnew}
\end{eqnarray}
From Eq.(\ref{eq:Bnew}),
the off diagonal elements of $B^2_{\rm new}$
are given as,
\begin{eqnarray}
B^2_{{\rm new} ij}= -2 A_0 \sum_{k=1}^2
O_{ik} (M_{R,k}-M_{R,3}) O_{jk} \quad (i \ne j).
\end{eqnarray}
The diagonal elements are dominated by the term proportional
to $B_0 M_{R,i}$,
\begin{eqnarray}
B^2_{{\rm new} ii}= B_0 D_i(M_{R,3})+ 2 A_0 (D_i-\sum_{k=1}^3
O_{ik}^2 M_{R,k}).
\end{eqnarray}
To write the off-diagonal elements of $B^2_{\rm new}$
explicitly, we introduce
the parametrization for the orthogonal matrix $O^T$ as,
\begin{eqnarray}
O^T= \begin{pmatrix} c^N_{13} c^N_{12} & c^N_{13} s^N_{12}
& s^N_{13} \\
-s^N_{23} s^N_{13} c^N_{12}-c^N_{23} s^N_{12}
& -s^N_{23} s^N_{13} s^N_{12} +c^N_{23} c^N_{12} & c^N_{13} s^N_{23} \\
 -c^N_{23} s^N_{13} c^N_{12}+ s^N_{23} s^N_{12}
& -c^N_{23} s^N_{13} s^N_{12} -s^N_{23}c^N_{12} &
c^N_{13} c^N_{23}
\end{pmatrix},
\end{eqnarray}
where $s^N_{ij}=\sin \theta^N_{ij}$ and
$c^N_{ij}=\cos \theta^N_{ij}$.
One can write
$B^2_{{\rm new},{ij}}$ $(i \ne j)$ using the angles as
\begin{eqnarray}
B^2_{{\rm new},{13}}&=&-2A_0 c^N_{13} \{(M_{R,1}-M_{R,2})c^N_{12}s^N_{13}
+(M_{R,2}-M_{R,3}) c^N_{23}(c^N_{23}c^N_{12}s^N_{13}-s^N_{23}
s^N_{12})\}, \nonumber
\\
B^2_{{\rm new},{23}}&=& -2A_0\{c^N_{13}s^N_{12}
s^N_{13}(M_{R,1}-M_{R,2})
+c^N_{13}c^N_{23}(s^N_{23}c^N_{12}+c^N_{23}s^N_{13} s^N_{12})
(M_{R,2}-M_{R,3})\},
\nonumber \\
B^2_{{\rm new},{12}}&=& -2A_0\{c^N_{12} s^N_{12}((M_{R,1}-M_{R,2})
-{s^N_{13}}^2 (M_{R,1}-M_{R,3}))  \nonumber \\
&+& s^N_{23} (s^N_{23}s^N_{12} c^N_{12}
(1+(s^N_{13})^2)-c^N_{23}s^N_{13}
 \cos 2\theta^N_{12})(M_{R,2}-M_{R,3}) \}.
\label{eq:Boff}
\end{eqnarray}
The angles $\theta^N_{ij}$ can be determined by the diagonalization
Eq.(\ref{eq:diag}).
When
$M_{R,3} \gg M_{R,2},  M_{R,1}$,
one can determine $s^N_{23}, s^N_{13}$ from the equation,
\begin{eqnarray}
&& \left(\begin{array}{ccc}
0& 0
& -2H_{13} t^{03}M_{R,3} \\
0 & 0
& -2H_{23}t^{03}M_{R,3}\\
-2H_{13} t^{03}M_{R,3} & -2H_{23}t^{03}M_{R,3}
& M_{R3}(M_{R3})\end{array} \right)
\begin{pmatrix} s^N_{13} \\
c^N_{13} s^N_{23} \\
 c^N_{13} c^N_{23} \end{pmatrix}
=D_3 \begin{pmatrix} s^N_{13} \\
                                 c^N_{13} s^N_{23} \\
                                 c^N_{13} c^N_{23} \end{pmatrix}.
\label{eq:app}
\end{eqnarray}
Then we find  that $D_{3} \simeq M_{R,3}$, and
$s^N_{13}$ and $s^N_{23}$
are given as
\begin{eqnarray}
&& s^N_{13}\simeq -2 H_{13} t^{03}, \nonumber \\
&& s^N_{23}\simeq -2 H_{23} t^{03},
\label{eq:sN}
\end{eqnarray}
where we ignore the corrections of the order of
$O(\frac{M_{R,i}}{M_{R,3}})$ $(i=1,2)$.
The determination of $s^N_{12}$ is more involved.
It is determined by diagonalizing
the following $2 \times 2$ matrix which can be obtained after
the largest eigenvalue state is decoupled from the $3 \times
3$ matrix in Eq.(\ref{eq:MMR3}).
\begin{eqnarray}
&&
\begin{pmatrix}
c^N_{12} & -s^N_{12} \\
s^N_{12} & c^N_{12}
\end{pmatrix}
\begin{pmatrix}
M_{R,1}-M_{R,3} s_{13}^2 &
-2H_{12} t^{03}M_{R,2} -M_{R,3} s_{13} s_{23} \\
-2 H_{12} t^{03}M_{R,2}-M_{R,3} s_{13} s_{23} &
M_{R,2}-M_{R,3} s_{23}^2
\end{pmatrix}
\begin{pmatrix}
c^N_{12} & s^N_{12} \\
-s^N_{12} & c^N_{12}
\end{pmatrix} \nonumber \\
&\simeq&\begin{pmatrix} D_1 & 0 \\
                        0 & D_2
                        \end{pmatrix}.
\end{eqnarray}
$s^N_{12}$ is approximately given by,
\begin{eqnarray}
s^N_{12}\simeq -2 H_{12} t^{03}-\frac{M_{R,3}}{M_{R,2}}
s^N_{13} s^N_{23},
\end{eqnarray}
where the following conditions are assumed to be satisfied,
\begin{eqnarray}
M_{R,2} > (s^N_{13})^2 M_{R,3} , (s^N_{23})^2 M_{R,3}.
\label{eq:cond}
\end{eqnarray}
Using the formulae given in Eq.(\ref{eq:sN}),
one can write the dominant terms for
the first two equations in Eq.(\ref{eq:Boff}),
\begin{eqnarray}
B^2_{{\rm new},{13}}&=&2A_0 M_{R,3} (c^N_{12} s^N_{13}-
s^N_{12}s^N_{23}),
\nonumber
\\
B^2_{{\rm new},{23}}&=& 2A_0 M_{R,3}(c^N_{12} s^N_{23}+
s^N_{12} s^N_{13}).
\end{eqnarray}
When $s^N_{12} \ll 1$, they are simplified as,
\begin{eqnarray}
B^2_{{\rm new},m 3}&=& 2 A_0 M_{R,3} s^N_{m 3} \nonumber \\
                  &=& -4 A_0 M_{R,3} H_{m 3} t^{03}, (m=1,2).
\label{eq:B2}
\end{eqnarray}
Eq.(\ref{eq:B2}) shows that in the diagonal basis
of $M_R$,
the off diagonal elements of $B^2_{m 3}$ is
given by the third term of Eq.(\ref{eq:B}) and
it is small compared with the large
diagonal element  $B^2_{{\rm new} 33} \sim B^0_N M_{R,3}$.

Next we show that
the variation of the threshold corrections
Eq.(\ref{eq:th_mlsq})
due to the change of the
basis is two loop effect and thus negligibly small.
When we change the basis as,
\begin{eqnarray}
(N_R)^c \rightarrow O^T {N_R}^c,
\end{eqnarray}
$Y_\nu$ in Eq.(\ref{eq:th_mlsq}) should be replaced by
\begin{eqnarray}
Y_\nu && \rightarrow Y_{\nu}O(M_{R,k})^T.
\label{replace}
\end{eqnarray}
Then the threshold correction is replaced by the
following equation,
\begin{eqnarray}
\delta^k_{\rm th} m^2_{\tilde{L}ij}= \frac{1}{8 \pi^2}
(Y_{\nu}^{\ast} O^T)_{ik} (O Y_{\nu}^T)_{kj} (m_0^2+ A_0 B^0_N).
\end{eqnarray}
Now let us examine how large
$\delta^k_{\rm th} m^2_{\tilde{L}ij}$ for the case $k=3$ could be
 after changing the basis as follows,
\begin{eqnarray}
\delta^3_{\rm th} m^2_{\tilde{L}ij}\simeq
\frac{m_0^2+ A_0 B^0_N}{8 \pi^2} \{Y^\ast_{\nu,i3} Y^T_{\nu,3j}+
Y^\ast_{\nu,i3} (Y^T_{\nu,2j} s^N_{23} + Y^T_{\nu,1j} s^N_{13})
+ (Y^\ast_{\nu,i2} s^N_{23} + Y^\ast_{\nu,i1}s^N_{13}) Y^T_{\nu,3j}
)\}. \nonumber \\
\end{eqnarray}
Since $s^N_{m3}= -(Y_\nu^T Y_{\nu}^\ast)_{m3}
\frac{1}{8 \pi^2}
t^{03}~~,(m=1,2)$, the variation $\delta^3_{\rm th} m^2_{\tilde{L}ij}$
is the two loop order and thus small
correction.

For the completeness, we check whether the
conditions Eq.(\ref{eq:cond}) and $s^N_{12} \ll 1$
which lead to Eq.(\ref{eq:B2})
are satisfied.
For  numerical estimation,
We take $M_{GUT}=10^{16}$(GeV), $M_{R,3}=10^{14}$(GeV),
$M_{R,2}=10^{12}$(GeV),and
$M_{R,1}=10^{10}$(GeV).
For $H_{ij}$, one may use the parametrization
Eq.(\ref{eq:Y}).
\begin{eqnarray}
H_{ij}=(Y_\nu^T Y_\nu^\ast)_{ij}=
\frac{1}{\langle H_2^0 \rangle^2}
\sqrt{M_{R,i} M_{R,j}} (R m_\nu R^\dagger)_{ij}.
\end{eqnarray}
When R is a real orthogonal matrix,
\begin{eqnarray}
|H_{ij}|< \frac{\sqrt{M_{R,i} M_{R,j}}}{v^2 \sin^2 \beta}
\times (|m_{\nu 3}-m_{\nu 1}|+ |m_{\nu 2}-m_{\nu 1}|).
\end{eqnarray}
Using $\tan \beta=5$ and $v=246$(GeV), we obtain
\begin{eqnarray}
|H_{12}| \le 1.0 \times 10^{-4},\nonumber \\
|H_{13}| \le 1.0 \times 10^{-3},\nonumber \\
|H_{23}| \le 1.0 \times 10^{-2}.
\end{eqnarray}
Since $2 t^{03} \simeq 0.06$, we obtain
\begin{eqnarray}
|s^N_{23}| \le  6 \times 10^{-4}, ~~~~~|s^N_{13}| \le 6 \times 10^{-5}.
\end{eqnarray}
Then Eq.(\ref{eq:cond}) is satisfied. We also note that
$s^N_{12}$
is as small as
\begin{eqnarray}
|s^N_{12}|\le  6 \times 10^{-6} + 3.6 \times 10^{-6} \sim
 1.0 \times
10^{-5}.
\end{eqnarray}


\end{document}